\documentclass[12pt,a4paper]{article}
\usepackage{graphicx}

\title{Reduction of the Bethe-Salpeter equation for the
scattering amplitude of the particles with spin 1 to system
of the integral equations for invariant functions }

\author{A. Yu. Loginov,\, V. N. Stibunov \\
{\it Nuclear Physics Institute} \\
{ \it at Tomsk Polytechnical University, Tomsk, Russia.}}
\date{}
\begin{document}
\maketitle
\begin{abstract}
Bethe-Salpeter equation for the massive particles with
spin 1 is considered. The scattering amplitude
decomposition of the particles with spin 1 by relativistic
tensors is derived.
The transformation coefficients from helicity amplitudes
to invariant functions is found.
The integral equations system for invariant functions
is obtained and  partial decomposition of this system is
performed.
Equivalent system of the integral equation for the partial
helicity amplitudes is presented.
\end{abstract}

\section{}
\hspace*{\parindent}

The Bethe-Salpeter equation was initially formulated in
quantum electrodynamics \cite{betsol, nambu} to describe the
bound two-body
states in the case where neither particle could be treated as
an external-field source. The applicability of this equation
was not, however, limited to the quantum electrodynamics
framework.
The Bethe-Salpeter equation can be formulated in other
renormalized models of quantum field theory, such as
$\phi^3$,\,$\phi^4$ and scalar electrodynamics \cite{itz}.
In addition, this equation is applied to describe strong
interactions
such as the $\pi N - $ and $N N - $ scattering
\cite{gerstein, schwartz, frohl},
the $ N N - N \Delta $ reactions \cite{br1, br2}, and other
processes that do
not fall within the framework of renormalized theories.
This equation is also used in describing the electromagnetic
interaction of hadrons, in particular, electron scattering on
a deuteron \cite{hummel1, hummel2}.  
The properties of scalar and vector mesons
within the constituent quark model can also be interpreted in
terms of the Bethe-Salpeter equation \cite{efrem,jain1,jain2}. 
A most frequent
application of the Bethe-Salpeter equation is for the spin-0
and spin- $\frac{1}{2}\;$ particle reaction amplitudes
\cite{henly1, henly2, lagae}.
This work considers the Bethe-Salpeter equation for the
scattering-reaction amplitude of the vector particles
$1^{-}+1^{-}\rightarrow 1^{-}+1^{-}$.  Mainly, it aims at
transforming the
Bethe-Salpeter equation for the spin-1 particle scattering
amplitude to a system of integral equations for invariant
functions.
Spin-1 particle scattering corresponds to e.g.
vector-meson interactions, vector-meson scattering on a
deuteron, or elastic deuteron scattering.
The paper is arranged as follows. Section 2 gives a general
expression for the $P-$ and $T-$ invariant amplitude of the
reaction  $1^{-}+1^{-}\rightarrow 1^{-}+1^{-}$. In Section 3,
the Bethe-Salpeter equation is reduced to a system of integral
equations for invariant
functions using the coefficients of the helicity amplitude
transformation to invariant functions. Section 4 presents the
partial extension of a system of $ 4D $ integral equations to
obtain a system of $ 2D $ integral equations. In Section 5, an
alternative system of $ 2D $ integral equations is given for
partial helicity amplitudes.   Appendix provides the
coefficients of helicity amplitude transformation to invariant
functions and the matrix of the system of $ 2D $ integral
equations for partial helicity amplitudes.

\section{}
\hspace*{\parindent}

Let us now consider the general structure of the
scattering-reaction amplitude for two spin-1 particles and
with a negative intrinsic parity $1^{-}+1^{-}\rightarrow 1^{-}+1^{-}$.
Let the initial particle momenta be $k_1$ and $q_1$, the final
momenta be $k_2$ and $q_2$,
the 4-vector amplitudes of the initial particle be $u$ and $v$,
and those for the final ones  $u^{'}$ and $v^{'}$.  To reveal
the amplitude symmetry with respect to the spatial reflection
and time inversion, it is convenient to make use of a
symmetrical and antisymmetrical combination of momenta:
\begin{eqnarray}
P=k_1+q_1=k_2+q_2,\quad p_1=\frac{1}{2}(k_1-q_1),
\quad p_2=\frac{1}{2}(k_2-q_2)\, .
\end{eqnarray}
The invariant variables $s,\, t $ and $u$ are expressed via
$P,\, p_1$, and $p_2$ in the following way:
\begin{eqnarray}
& & s=P^2,\quad t=(k_1-k_2)^2=(p_1-p_2)^2,
\quad u=(k_1-q_2)^2=(p_1+p_2)^2, \nonumber   \\
& &  s+t+u=2m_1^2+2m_2^2 \, .
\end{eqnarray}
Let us assume the interaction between the particles to be
$P-$ and $T-$ invariant. A general expression for the reaction
amplitude with arbitrary spins will then be:
\begin{eqnarray}
T(p_2,p_1;P)=\sum_{i} f_i(s,t)R^{i}(p_2,p_1;P),
\end{eqnarray}
where $f_i(s,t)$ are the invariant functions depending on the
initial- and final-particle 4-momenta via the invariant
variables $s$ and $t$ only, and $R^{i}(p_2,p_1;P)$ are the
invariant combinations  made of 4-momenta and wave functions
of all particles participating in the reaction. To construct
the reaction amplitude, we need to determine the number of
independent invartiant functions entering Eq. (3).
It is equal to the number of independent helicity reaction
amplitudes taking into account the $P-$ and $T-$ invariance.
The total number of the helicity amplitudes of the reaction
$1^{-}+1^{-}\rightarrow 1^{-}+1^{-}$ is equal to $(2s_1+1)(2s_2+1)
(2s_3+1)(2s_4+1)= 81$.
These amplitudes are related in a manner determined by
the $P-$ invariant interactions
\begin{eqnarray}
T(\lambda_3,\lambda_4;\;\lambda_1,\lambda_2)=
\eta(-1)^{(\lambda_1-\lambda_2)-(\lambda_3-\lambda_4)}
T(-\lambda_3,-\lambda_4;\;-\lambda_1,-\lambda_2)\, ,
\end{eqnarray}
where $\eta=\eta_1 \eta_2 \eta_3 \eta_4 (-1)^{s_3+s_4-s_1-s_2},\;$
$\eta_{i},$ and $ s_{i} \, - $ are the intrinsic parity and
spin of the particles.
It is clear that the number of
independent helicity amplitudes decreases down to 41.
Further restrictions on the number of independent spiral
amplitudes follow from the $T-$ invariance of the interaction.
For elastic processes, the $T-$ invariance gives rise to the
following relations between the helicity amplitudes:
\begin{eqnarray}
T(\lambda_3,\lambda_4;\;\lambda_1,\lambda_2)=
(-1)^{(\lambda_1-\lambda_2)-(\lambda_3-\lambda_4)}
T(\lambda_1,\lambda_2;\;\lambda_3,\lambda_4)\, .
\end{eqnarray}
In this case, the number of helicity amplitudes is decreased
to 25. The following helicity amplitudes can be selected as
independent ones:
\begin{eqnarray}
& & T(1,1;1,1),\,T(1,1;1,0),\,T(1,1;1,-1),\,T(1,1;0,1),\,
T(1,1;0,0),\,T(1,1;0,-1),\nonumber   \\
& & T(1,1;-1,1),\,T(1,1;-1,0),\,T(1,1;-1,-1),\,
T(1,0;1,0),\,T(1,0;1,-1),\,T(1,0;0,1),\nonumber   \\
& & T(1,0;0,0),\,T(1,0;0,-1),\,T(1,0;-1,1),\,T(1,0;-1,0),\,
T(1,-1;1,-1),\,T(1,-1;0,1),\nonumber   \\
& & T(1,-1;0,0),\,T(1,-1;0,-1),\,T(1,-1;-1,1),\,
T(0,1;0,1),\,T(0,1;0,0),\,T(0,1;0,-1),\nonumber   \\
& &T(0,0;0,0)\, .
\end{eqnarray}

Now, we have to identify 25 independent invariant spin
combinations $R^{i}(p_2,p_1;P)$ entering into Eq.(3).
Since $s,\, t$ and $u$ are $P-$ and $T-$ invariant,
then $f_i(s,t)$ is also $P-$ and $T-$ invariant.
The values $R^{i}(p_2,p_1;P)=u^{' * \mu} v^{' * \nu}
R_{\mu \nu \alpha \beta }^{i} u^{\alpha} v^{\beta} $,
where $ u', \, v', \, u, \, v $ are the 4-vector
polarizations of final and initial particles, should,
therefore, be also $P-$ and $T-$ invariant.
The following set of the 4th rank 4-tensors satisfying
the $P-$ and $T-$ invariance could be selected as:
\begin{eqnarray}
& & R_{\mu \nu \alpha \beta }^{1}=p_{1\mu} p_{1\nu}p_{2\alpha} p_{2\beta},\;
R_{\mu \nu \alpha \beta }^{2}=p_{1\mu} P_{\nu}p_{2\alpha} p_{2\beta}+
p_{1\mu} p_{1\nu}p_{2\alpha} P_{\beta},\;
R_{\mu \nu \alpha \beta }^{3}=P_{\mu} p_{1 \nu}p_{2\alpha} p_{2\beta}+
p_{1\mu} p_{1\nu}P_{\alpha} p_{2\beta},\;\nonumber  \\
& &R_{\mu \nu \alpha \beta }^{4}=P_{\mu} P_{\nu}p_{2\alpha} p_{2\beta}+
p_{1\mu} p_{1\nu}P_{\alpha} P_{\beta}, \;
R_{\mu \nu \alpha \beta }^{5}=p_{1\mu}P_{\nu}p_{2\alpha}P_{\beta},\;
R_{\mu \nu \alpha \beta }^{6}=p_{1\mu}P_{\nu}P_{\alpha}p_{2\beta}+
P_{\mu}p_{1\nu}p_{2\alpha}P_{\beta}, \;\nonumber  \\
& & R_{\mu \nu \alpha \beta }^{7}=P_{\mu}P_{\nu}p_{2\alpha}P_{\beta}+
p_{1\mu}P_{\nu}P_{\alpha}P_{\beta}, \;R_{\mu \nu \alpha \beta }^{8}=
P_{\mu}p_{1\nu}P_{\alpha}p_{2\beta}, \; R_{\mu \nu \alpha \beta }^{9}
=P_{\mu}P_{\nu}P_{\alpha}p_{2\beta}+P_{\mu}p_{1\nu}P_{\alpha}P_{\beta}, \;\nonumber  \\
& & R_{\mu \nu \alpha \beta }^{10}=P_{\mu}P_{\nu}P_{\alpha}P_{\beta}, \;
R_{\mu \nu \alpha \beta }^{11}=g_{\mu \nu}p_{2\alpha}p_{2\beta}+
p_{1\mu} p_{1\nu}g_{\alpha \beta}, \;
R_{\mu \nu \alpha \beta }^{12}=g_{\mu \nu}p_{2\alpha}P_{\beta}+
p_{1\mu} P_{\nu}g_{\alpha \beta}, \;\nonumber  \\
& & R_{\mu \nu \alpha \beta }^{13}=g_{\mu \nu}P_{\alpha}p_{2\beta}+
P_{\mu}p_{1\nu}g_{\alpha \beta}, \;
R_{\mu \nu \alpha \beta }^{14}=g_{\mu \nu}P_{\alpha}P_{\beta}+
P_{\mu} P_{\nu}g_{\alpha \beta}, \;R_{\mu \nu \alpha \beta }^{15}=
p_{1\mu}g_{\nu \beta}p_{2\alpha}, \;\nonumber  \\
& & R_{\mu \nu \alpha \beta }^{16}=P_{\mu}g_{\nu \beta}p_{2\alpha}+
p_{1\mu}g_{\nu \beta}P_{\alpha}, \;
R_{\mu \nu \alpha \beta }^{17}=P_{\mu}g_{\nu \beta}P_{\alpha}, \;
R_{\mu \nu \alpha \beta }^{18}=g_{\mu \beta}p_{1\nu}p_{2\alpha}+
p_{1\mu}g_{\nu \alpha}p_{2\beta}, \;\nonumber  \\
& & R_{\mu \nu \alpha \beta }^{19}=g_{\mu \beta}P_{\nu}p_{2\alpha}+
p_{1\mu}g_{\nu \alpha}P_{\beta}, \;
R_{\mu \nu \alpha \beta }^{20}=g_{\mu \beta}p_{1\nu}P_{\alpha}+
P_{\mu}g_{\nu \alpha}p_{2\beta}, \;
R_{\mu \nu \alpha \beta }^{21}=g_{\mu \beta}P_{\nu}P_{\alpha}+
P_{\mu}g_{\nu \alpha}p_{\beta}, \;\nonumber  \\
& &R_{\mu \nu \alpha \beta }^{22}=g_{\mu \alpha}p_{1\nu}p_{2\beta}, \;
R_{\mu \nu \alpha \beta }^{23}=g_{\mu \alpha}P_{\nu}p_{2\beta}+
g_{\mu \alpha}p_{1\nu}P_{\beta}, \;R_{\mu \nu \alpha \beta }^{24}=
g_{\mu \alpha}P_{\nu}P_{\beta}, \;\nonumber  \\
& & R_{\mu \nu \alpha \beta }^{25}=g_{\mu \alpha}g_{\nu \beta}\, .
\end{eqnarray}
From Eqs.(4) and (5), it is evident that the relations of the
$P-$ and $T-$ invariance between the helicity amplitudes for
the elastic scattering reaction between the spin 1 - particles
coincide, provided the product of their internal
parities $\eta_1 \eta_2 \eta_3 \eta_4 $ equals unity.
Note that the number of independent helicity amplitudes also
coincides. Hence, the 4-tensor set (Eq. 7) can be used to construct
amplitudes for the reactions  $1^{+}+1^{+}\rightarrow 1^{+}+1^{+},\quad
1^{+}+1^{+}\rightarrow 1^{-}+1^{-},\quad
1^{+}+1^{-}\rightarrow 1^{+}+1^{-}$.

The 4-tensor set (Eq.7) is not the only possibility.
Consider now an invariant spin combination
$R^{25}(p_2,p_1;P)= u^{'*}\cdot u \,v^{'*}\cdot v$.
A question arises as to why there are no $u^{'*}\cdot v^{'*}\,u\cdot v $
or $ u^{'*}\cdot v\,v^{'*}\cdot u \;$
spin combinations among the invariant spin combinations
(Eq.8) similar to $R^{25}(p_2,p_1;P)$.
The answer to this question lies in the fact that these spin
combinations are not independent, which can be proved using
Gram's determinant.
It is well known that in a $ 4D $ space any five 4-vectors
are linearly related.
In our case this implies the Gram's
determinant is zero:
\[G \left( \begin{array}{ccccc}
u, & u^{'*}, & P, & p_1, & p_2 \\
v, & v^{'*}, & P, & p_1, & p_2
\end{array} \right)= 0\, , \]
which contains the polarization vectors of all the particles
participating in the reaction, and all the 4-momenta entering
the expression for the 4-tensors
$R_{\mu \nu \alpha \beta }^{i}$.
Now, one has to determine the number of independent relations
conditioned by Gram's determinants.
Due to the symmetry properties of Gram's determinants 
\cite{bykling},
any vector permutations within the upper or lower rows are
possible - this may result in a change of their common sign
only. It is the vector permutation from the upper or lower
rows in Gram's determinant that could give rise to essentially new
relations.
It should  be  noted  that  the  permutation   between   the
 $P,\,p_1,\,p_2$  vectors from the upper and lower rows cannot
result in new couplings either, since for any such permutation,
there will be two similar 4-vectors in  the  upper  and  lower
rows, which  would  result  in  a  identically  zero  Gram's
determinant.
Thus, the remaining permutations are those in the group
$  \begin{array}{cc}
u, & u^{'*}\\
v, & v^{'*}
\end{array}  $.
With regard to the symmetry properties of Gram's
determinant, the following permutations occur:
 \begin{equation}
\begin{array}{cc}
v, & u^{'*}\\
u, & v^{'*}
\end{array},\qquad
\begin{array}{cc}
u, & v\\
u^{'*}, & v^{'*}
\end{array}\, .
\end{equation}

Thus, three Gram's determinants are equal to zero:
\begin{eqnarray}
& & G \left( \begin{array}{ccccc}
u, & u^{'*}, & P, & p_1, & p_2 \\
v, & v^{'*}, & P, & p_1, & p_2
\end{array} \right)= 0, \;
G \left( \begin{array}{ccccc}
v, & u^{'*}, & P, & p_1, & p_2 \\
u, & v^{'*}, & P, & p_1, & p_2
\end{array} \right)= 0,  \nonumber \\
& & G \left( \begin{array}{ccccc}
u, & v, & P, & p_1, & p_2 \\
u^{'*}, & v^{'*}, & P, & p_1, & p_2
\end{array} \right)= 0\, .
\end{eqnarray}
Removing brackets, we obtain a homogenous system of three
linear equations. This system is, however, of the second rank,
since the third linear equation appears to be equal to the
difference between the first and second equations.
There are two linearly independent relations between the
invariant spin combinations, which allows expressing
$u^{'*}\cdot v^{'*}\,u\cdot v$ and
$u^{'*}\cdot v\,v^{'*}\cdot u$
in terms of $R^{1}-R^{25}$:
\begin{eqnarray}
& &u^{'*}\cdot v^{'*}\,u\cdot v =
 \frac{4\,{R1}}{t\,\left( -4\,m^2 + s + t \right) } +
 \frac{{R4}\,\left( 4\,m^2 - s \right) }
   {s\,t\,\left( -4\,m^2 + s + t \right) }-
\frac{{R5}\,\left( -4\,m^2 + s + 2\,t \right) }
   {s\,t\,\left( -4\,m^2 + s + t \right) } + \nonumber \\
& & +\frac{{R6}}{t\,\left( -4\,m^2 + s + t \right) }+
    \frac{{R7}}{-4\,m^2\,s + s^2 + s\,t}-
  \frac{{R8}\,\left( -4\,m^2 + s + 2\,t \right) }
   {s\,t\,\left( -4\,m^2 + s + t \right) }
   - \frac{{R9}}
   {s\,\left( -4\,m^2 + s + t \right) }+\nonumber \\
& & + \frac{{R10}\,\left( s - 4\,t \right) }
   {4\,s\,t\,\left( -4\,m^2 + s + t \right) }-
   \frac{{R11}\,\left( 4\,m^2 - s \right) }
 {t\,\left( -4\,m^2 + s + t \right) }
-\frac{{R12}\,\left( -4\,m^2 + s + 2\,t \right) }
   {2\,t\,\left( -4\,m^2 + s + t \right) }+
  \frac{{R13}\,\left( -4\,m^2 + s + 2\,t \right) }
   {2\,t\,\left( -4\,m^2 + s + t \right) } + \nonumber \\
& & + \frac{{R14}\,\left( -s^2 + 4\,m^2\,\left( s - 4\,t \right)  +
4\,s\,t + 4\,t^2 \right) }{4\,s\,t\,\left( -4\,m^2 + s + t \right) }
+\frac{{R15}\,\left( -4\,m^2 + s + 2\,t \right) }
   {t\,\left( -4\,m^2 + s + t \right) }
-\frac{{R16}\,\left( 4\,m^2 - s \right) }
   {2\,t\,\left( -4\,m^2 + s + t \right) }+ \nonumber \\
& & + \frac{{R17}\,\left( s^2 - 4\,m^2\,\left( s - 4\,t \right)  -
    2\,s\,t - 4\,t^2 \right) }{4\,s\,t\,\left( -4\,m^2 + s +
    t \right) }+ \frac{{R22}\,\left( -4\,m^2 + s + 2\,t \right) }
   {t\,\left( -4\,m^2 + s + t \right) } -
  \frac{{R23}\,\left( -4\,m^2 + s \right) }
   {2\,t\,\left( -4\,m^2 + s + t \right) }+ \nonumber \\
& & + \frac{{R24}\,\left( s^2 - 4\,m^2\,\left( s - 4\,t \right)  -
2\,s\,t - 4\,t^2 \right) }{4\,s\,t\,\left( -4\,m^2 + s + t \right) }+
{R25}\, .
\end{eqnarray}
\begin{eqnarray}
& & u^{'*}\cdot v\,v^{'*}\cdot u=-
\frac{2\, {R5}\,\left( -2\,m^2 + t \right) }
{s\,t\,\left( -4\,m^2 + s + t \right) }+
\frac{2\, {R6}\,\left( -2\,m^2 + s + t \right) }
{s\,t\,\left( -4\,m^2 + s + t \right) }+
\frac{{R7}\,\left( 4\,m^2 - s \right) }
{s\,t\,\left( -4\,m^2 + s + t \right) }-\nonumber \\
& & - \frac{2\, {R8}\,\left( -2\,m^2 + t \right) }
{s\,t\,\left( -4\,m^2 + s + t \right) }-
\frac{{R9}\,\left( 4\,m^2 - s \right) }
{s\,t\,\left( -4\,m^2 + s + t \right) }-
 \frac{{R10}\,\left( -4\,m^2 + s + 2\,t \right) }
{s\,t\,\left( -4\,m^2 + s + t \right) }+
\frac{{R15}\,\left( -4\,m^2 + s + 2\,t \right) }
{t\,\left( -4\,m^2 + s + t \right) }-\nonumber \\
& & - \frac{{R16}\,\left( 4\,m^2 - s \right) }
{2\,t\,\left( -4\,m^2 + s + t \right) }
+\frac{{R17}\,\left( s^2 - 4\,m^2\,\left( s - 4\,t \right)  -
2\,s\,t - 4\,t^2 \right) }{4\,s\,t\,\left( -4\,m^2 + s + t \right) } \
-\frac{{R18}\,\left( -4\,m^2 + s + 2\,t \right) }
{t\,\left( -4\,m^2 + s + t \right) } -\nonumber \\
& & - \frac{{R19}\,\left( 4\,m^2 - s \right) }
{2\,t\,\left( -4\,m^2 + s + t \right) }
-  \frac{{R20}\,\left( -4\,m^2 + s \right) }
{2\,t\,\left( -4\,m^2 + s + t \right) }+
\frac{{R21}\,\left( s^2 + 6\,s\,t + 4\,t^2 -
4\,m^2\,\left( s + 4\,t \right)  \right) }{4\,s\,t\,
\left( -4\,m^2 + s + t \right) } + \\
& & + \frac{{R22}\,\left( -4\,m^2 + s + 2\,t \right) }
{t\,\left( -4\,m^2 + s + t \right) }-
\frac{{R23}\,\left( -4\,m^2 + s \right) }
{2\,t\,\left( -4\,m^2 + s + t \right) }+
\frac{{R24}\,\left( s^2 - 4\,m^2\,\left( s - 4\,t \right)  -
2\,s\,t - 4\,t^2 \right) }
{4\,s\,t\,\left( -4\,m^2 + s + t \right) } \
+{R25}\, .\nonumber
\end{eqnarray}
When constructing the amplitude Eq.(3), use can be made
of the spin combinations $u^{'*}\cdot v^{'*}\,u\cdot v,\quad
u^{'*}\cdot v\,v^{'*}\cdot u$  instead of any other  two spin
combinations $R^{i}$ from Eqs.(10) and (11).

Let us write a general expression for the $P-$ and $T-$
invariant   helicity   amplitude    of   the   reaction
$1^{-}+1^{-}\rightarrow 1^{-}+1^{-}$:
\begin{eqnarray}
& &T_{\lambda_3\,\lambda_4,\, \lambda_1\,\lambda_2}(p_2,p_1;P)=
u_{\lambda_3}^{'*}\cdot {p_1}\,
v_{\lambda_4}^{'*}\cdot {p_1}\,
{u_{{{\lambda }_1}}}\cdot {p_2}\,
{v_{{{\lambda }_2}}}\cdot {p_2}\,{f_1} \
+ \nonumber  \\ & & + ( u_{\lambda_3}^{'*}\cdot {p_1}\,
v_{\lambda_4}^{'*}\cdot {p_1}\,
{u_{{{\lambda }_1}}}\cdot {p_2}\,
{v_{{{\lambda }_2}}}\cdot P +
u_{\lambda_3}^{'*}\cdot {p_1}\,
v_{\lambda_4}^{'*}\cdot P\,
{u_{{{\lambda }_1}}}\cdot {p_2}\,
{v_{{{\lambda }_2}}}\cdot {p_2} ) \,{f_2} +\nonumber  \\
& & + ( u_{\lambda_3}^{'*}\cdot {p_1}\,
v_{\lambda_4}^{'*}\cdot {p_1}\,
{u_{{{\lambda }_1}}}\cdot P\,{v_{{{\lambda }_2}}}\cdot {p_2} +
u_{\lambda_3}^{'*}\cdot P\,
v_{\lambda_4}^{'*}\cdot {p_1}\,
{u_{{{\lambda }_1}}}\cdot {p_2}\,
{v_{{{\lambda }_2}}}\cdot {p_2} \ ) \,{f_3} + \nonumber  \\
& & + (  u_{\lambda_3}^{'*}\cdot {p_1}\,
v_{\lambda_4}^{'*}\cdot {p_1}\,
{u_{{{\lambda }_1}}}\cdot P\,{v_{{{\lambda }_2}}}\cdot P +
u_{\lambda_3}^{'*}\cdot P\,
v_{\lambda_4}^{'*}\cdot P\,
{u_{{{\lambda }_1}}}\cdot {p_2}\,
{v_{{{\lambda }_2}}}\cdot {p_2} ) \,{f_4} +\nonumber  \\
& & + u_{\lambda_3}^{'*}\cdot {p_1}\,
v_{\lambda_4}^{'*}\cdot P\,
{u_{{{\lambda }_1}}}\cdot {p_2}\,
{v_{{{\lambda }_2}}}\cdot P\,{f_5} +
( u_{\lambda_3}^{'*}\cdot P\,
v_{\lambda_4}^{'*}\cdot {p_1}\,
{u_{{{\lambda }_1}}}\cdot {p_2}\,
{v_{{{\lambda }_2}}}\cdot P +\nonumber  \\ & & +
u_{\lambda_3}^{'*}\cdot {p_1}\,
v_{\lambda_4}^{'*}\cdot P
\,{u_{{{\lambda }_1}}}\cdot P\,
{v_{{{\lambda }_2}}}\cdot {p_2} ) \,{f_6} +
( u_{\lambda_3}^{'*}\cdot {p_1}\,
v_{\lambda_4}^{'*}\cdot P\,
{u_{{{\lambda }_1}}}\cdot P\,
{v_{{{\lambda }_2}}}\cdot P +\nonumber  \\ & & +
u_{\lambda_3}^{'*}\cdot P\,
v_{\lambda_4}^{'*}\cdot P\,
{u_{{{\lambda }_1}}}\cdot {p_2}\,{v_{{{\lambda }_2}}}
\cdot P )\,{f_7} +
u_{\lambda_3}^{'*}\cdot P\,
v_{\lambda_4}^{'*}\cdot {p_1}\,
{u_{{{\lambda }_1}}}\cdot P\,
{v_{{{\lambda }_2}}}\cdot {p_2}\,{f_8} +\nonumber  \\
& & + ( u_{\lambda_3}^{'*}\cdot P\,
v_{\lambda_4}^{'*}\cdot {p_1}\,
{u_{{{\lambda }_1}}}\cdot P\,{v_{{{\lambda }_2}}}\cdot P +
u_{\lambda_3}^{'*}\cdot P\,
v_{\lambda_4}^{'*}\cdot P\,
{u_{{{\lambda }_1}}}\cdot P\,
{v_{{{\lambda }_2}}}\cdot {p_2} ) \,{f_9} +\nonumber  \\ & & +
u_{\lambda_3}^{'*}\cdot P\,
v_{\lambda_4}^{'*}\cdot P\,{u_{{{\lambda }_1}}}
\cdot P\,{v_{{{\lambda }_2}}}\cdot P\,{f_{10}} +
( u_{\lambda_3}^{'*}\cdot {p_1}\,
v_{\lambda_4}^{'*}\cdot {p_1}\,
{u_{{{\lambda }_1}}}\cdot {v_{{{\lambda }_2}}} +
u_{\lambda_3}^{'*}\cdot v_{\lambda_4}^{'*}\,
{u_{{{\lambda }_1}}}\cdot {p_2}\,
{v_{{{\lambda }_2}}}\cdot {p_2} ) \,{f_{11}} +\nonumber  \\ & &
 + ( u_{\lambda_3}^{'*}\cdot {p_1}\,
v_{\lambda_4}^{'*}\cdot P\,
{u_{{{\lambda }_1}}}\cdot {v_{{{\lambda }_2}}} +
u_{\lambda_3}^{'*}\cdot v_{\lambda_4}^{'*}\,
{u_{{{\lambda }_1}}}\cdot {p_2}\,
{v_{{{\lambda }_2}}}\cdot P ) \,{f_{12}} +\nonumber  \\ & & +
 ( u_{\lambda_3}^{'*}\cdot P\,
v_{\lambda_4}^{'*}\cdot {p_1}\,
{u_{{{\lambda }_1}}}\cdot {v_{{{\lambda }_2}}} +
u_{\lambda_3}^{'*}\cdot v_{\lambda_4}^{'*}\,
{u_{{{\lambda }_1}}}\cdot P\,{v_{{{\lambda }_2}}}
\cdot {p_2} ) \,{f_{13}} +\nonumber  \\
& & + ( u_{\lambda_3}^{'*}\cdot P\,v_{\lambda_4}^{'*}\cdot P\,
{u_{{{\lambda }_1}}}\cdot {v_{{{\lambda }_2}}} +
u_{\lambda_3}^{'*}\cdot v_{\lambda_4}^{'*}\,
{u_{{{\lambda }_1}}}\cdot P\,{v_{{{\lambda }_2}}}\cdot P )
\,{f_{14}} + u_{\lambda_3}^{'*}\cdot {p_1}\,
v_{\lambda_4}^{'*}\cdot {v_{{{\lambda }_2}}}\,
{u_{{{\lambda }_1}}}\cdot {p_2}\,{f_{15}} +\nonumber  \\
& & + ( u_{\lambda_3}^{'*}\cdot {p_1}\,
v_{\lambda_4}^{'*}\cdot {v_{{{\lambda }_2}}}\,
{u_{{{\lambda }_1}}}\cdot P +
u_{\lambda_3}^{'*}\cdot P\,
v_{\lambda_4}^{'*}\cdot {v_{{{\lambda }_2}}}\,
{u_{{{\lambda }_1}}}\cdot {p_2} ) \,{f_{16}} +
u_{\lambda_3}^{'*}\cdot P\,
v_{\lambda_4}^{'*}\cdot {v_{{{\lambda }_2}}}\,
{u_{{{\lambda }_1}}}\cdot P\,{f_{17}} +\nonumber  \\
& & + (  u_{\lambda_3}^{'*}\cdot {v_{{{\lambda }_2}}}\,
v_{\lambda_4}^{'*}\cdot {p_1}\,
{u_{{{\lambda }_1}}}\cdot {p_2} +
u_{\lambda_3}^{'*}\cdot {p_1}\,
v_{\lambda_4}^{'*}\cdot {u_{{{\lambda }_1}}}\,
{v_{{{\lambda }_2}}}\cdot {p_2} ) \,{f_{18}} +
(  u_{\lambda_3}^{'*}\cdot {v_{{{\lambda }_2}}}\,
v_{\lambda_4}^{'*}\cdot P\,
{u_{{{\lambda }_1}}}\cdot {p_2} +\nonumber  \\
& & + u_{\lambda_3}^{'*}\cdot {p_1}\,
v_{\lambda_4}^{'*}\cdot {u_{{{\lambda }_1}}}\,
{v_{{{\lambda }_2}}}\cdot P ) \,{f_{19}} +
(  u_{\lambda_3}^{'*}\cdot {v_{{{\lambda }_2}}}\,
v_{\lambda_4}^{'*}\cdot {p_1}\,
{u_{{{\lambda }_1}}}\cdot P +
u_{\lambda_3}^{'*}\cdot P\,
v_{\lambda_4}^{'*}\cdot {u_{{{\lambda }_1}}}\,
{v_{{{\lambda }_2}}}\cdot {p_2} ) \,{f_{20}} +\nonumber  \\
& & + ( u_{\lambda_3}^{'*}\cdot {v_{{{\lambda }_2}}}\,
v_{\lambda_4}^{'*}\cdot P\,
{u_{{{\lambda }_1}}}\cdot P +
u_{\lambda_3}^{'*}\cdot P\,
v_{\lambda_4}^{'*}\cdot {u_{{{\lambda }_1}}}\,
{v_{{{\lambda }_2}}}\cdot P ) \,{f_{21}} +
u_{\lambda_3}^{'*}\cdot {u_{{{\lambda }_1}}}\,
v_{\lambda_4}^{'*}\cdot {p_1}\,
{v_{{{\lambda }_2}}}\cdot {p_2}\,{f_{22}} +\nonumber  \\
& & + ( u_{\lambda_3}^{'*}\cdot {u_{{{\lambda }_1}}}\,
v_{\lambda_4}^{'*}\cdot {p_1}\,
{v_{{{\lambda }_2}}}\cdot P +
u_{\lambda_3}^{'*}\cdot {u_{{{\lambda }_1}}}\,
v_{\lambda_4}^{'*}\cdot P\,
{v_{{{\lambda }_2}}}\cdot {p_2}) \,{f_{23}} +
u_{\lambda_3}^{'*}\cdot
{u_{{{\lambda }_1}}}\,v_{\lambda_4}^{'*}\cdot P\,
{v_{{{\lambda }_2}}}\cdot P\,{f_{24}} +\nonumber  \\
& & + u_{\lambda_3}^{'*}\cdot {u_{{{\lambda }_1}}}\,
v_{\lambda_4}^{'*}\cdot {v_{{{\lambda }_2}}}\,{f_{25}}\, ,
\end{eqnarray}
where      $\lambda_3, \lambda_4$ and $\lambda_1, \lambda_2$
are the helicities  of the vector particles in their initial
and final states.

\section{}
\hspace*{\parindent}

The Bethe - Salpeter equation is a relativistic relation
for the two-body Green function $ G(x_1',x_2'; x_1, x_2)$:
\begin{eqnarray}
G(x_1',x_2'; x_1, x_2)=I(x_1',x_2'; x_1, x_2)+
\int K(x_1',x_2'; x_3, x_4) G( x_3, x_4; x_1, x_2)
d^4 x_3 d^4 x_4 \, ,
\end{eqnarray}
where $x_1, x_2, x_1', x_2'$  are the initial and final
$4D$-coordinates of a particle.
\begin{figure}
\begin{center}
\includegraphics[width=12cm]{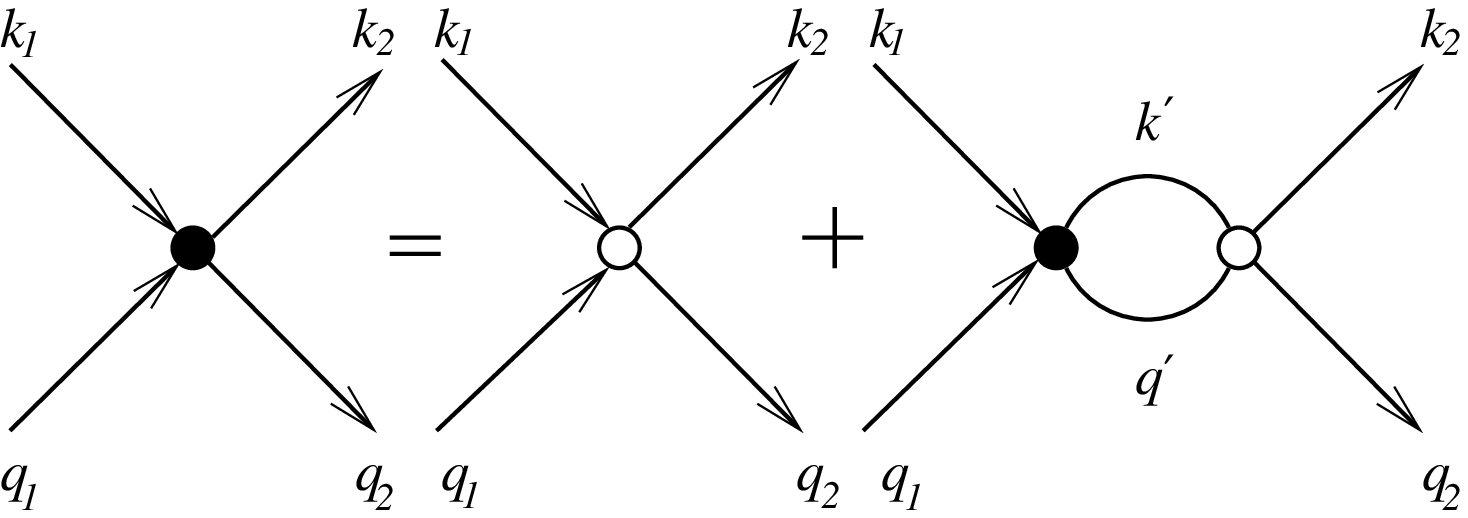}
\end{center}
\parbox[t]{16cm}{
\begin{center}
\Large{\textbf {Fig.1}}
\end{center}}
\end{figure}
Thus, the Bethe-Salpeter equation relates the total Green
function for two particles $ G(x_1',x_2'; x_1, x_2),$
representing a sum of all
the Feinman diagrams (Fig. 1, left) to a certain
topologically derived part of this function
$I(x_1',x_2'; x_1, x_2)$
(Fig. 1 the first term in the right side)  ,
which is a sum of all two-body irreducible diagrams in
the $s- $ channel , i.e. , the diagrams that could not be
split into two linked parts containing points $x_1, x_2$ and
$x_1', x_2' $
by breaking two lines  in the direction of the $s - $ channel.
The kernel $K$ of the Bethe - Salpeter equation is explicitly
expressed through the sum of two-body irreducible diagrams $I$
and single-body Green functions for the scattering particles.
In Fig. 1, the second term in the right side corresponds to
the integral term in Eq.(13).
The kernel of the Bethe - Salpeter equation is based on
Lagrangian particle interaction.
Thus, Eq.(13)
provides an expression for the total Feinman diagrams via
their two-body irreducible part. Since the kernel and the
inhomogeneous term of the equation are derived using the
approaches of the perturbation theory, they may only be
approximated. That is why it is an approximated equation
resulting, when one reduces the manipulations in the sum of
irreducible diagrams to the lowest perturbation theory orders,
where between the two interacting particles there is a single
quantum exchange, which is often taken for the Bethe-Salpeter
equation. This approximation is called a ladder approach.
In a momentum space, the Bethe - Salpeter equation (13) has
the form of an integral equation for the two-body scattering
amplitude $ T(p_2, p_1; P)$:
\begin{eqnarray}
 T(p_2, p_1; P)=I(p_2, p_1; P) +
\int K(p_2, p';P)T(p', p_1; P)\frac{d^4 p'}{(2 \pi)^4} \, ,
\end{eqnarray}
where $I$ and $K$ are the inhomogeneous term and the kernel
of the equation. Equation (14) for the vector-particle
scattering amplitude can be written as follows:
\begin{eqnarray}
& & T_{\mu \nu  \alpha \beta }({p_2},{p_1};P) =
I_{\mu \nu  \alpha \beta }({p_2},{p_1};P) +
\int I_{\mu \nu  \epsilon \eta }({p_2},{p'};P)
 G^{\epsilon \,\gamma }\,\left( \frac{P}{2} + p' \right)
 G^{\eta \,\delta }\,\left( \frac{P}{2} - p' \right)\times \nonumber \\
& & \times T_{\gamma \delta  \alpha \beta }({p'},{p_1};P)\frac{d^4 p'}
{(2 \pi)^4} \,  ,
\end{eqnarray}
where $I$ is the sum of two-body irreducible diagrams and  $G$
are the vector particle propagators. In Eq. (15), the 4-vector
particle polarizations are omitted, and the sequence of the
4-indices corresponds to the motion opposite the lines shown
in Fig. 1.
Let us express the amplitude $T_{\mu \nu  \alpha \beta }$ and
the sum of two-body irreducible diagrams
$I_{\mu \nu  \alpha \beta }$ as an expansion in tensors
$\;R_{\mu \nu \alpha \beta }^{i}$:
\begin{eqnarray}
T_{\mu \nu  \alpha \beta }({p_2},{p_1};P) =
\sum_{i = 1}^{25}{f_i}({p_2},{p_1};P)\,
R_{\mu \nu \alpha \beta }^{i} ({p_2},{p_1};P),
\end{eqnarray}
\begin{eqnarray}
I_{\mu \nu  \alpha \beta }({p_2},{p_1};P) =
\sum_{i = 1}^{25}{g_i}({p_2},{p_1};P)\,
R_{\mu \nu \alpha \beta }^{i} ({p_2},{p_1};P).
\end{eqnarray}
Now, proceed from the 4-tensors $T_{\mu \nu  \alpha \beta }$
to the helicity amplitudes:
\begin{eqnarray}
T_{\lambda_3 \,\lambda _4,
\lambda_1\,\lambda_2}(p_2,p_1; P) =
{v^{*}}_{\lambda_3\,\lambda_4}^{\mu \,\nu } (p_2;P)\, T_{\mu \nu \alpha \beta }(p_2,p_1;P)\,
v_{\lambda_1 \,\lambda_2}^{\alpha \,\beta } (p_1;P) \, ,
\end{eqnarray}
where the ${v^*}_{\lambda_3\,\lambda_4}^{\mu \,\nu } (p_2;P)$
and      $v_{\lambda_1\,\lambda_2}^{\alpha \,\beta } (p_1;P)$
are the products of the helicity 4-vector
polarizations of the final and initial particles
\begin{eqnarray}
& & {v^*}_{\lambda_3\,\lambda_4}^{\mu \,\nu } (p_2;P)=
u^{'* \,\mu } \left( \frac{P}{2} + {p_2},\lambda_3 \right)
\,v^{'*\,\nu } \left(\frac{P}{2} - {p_2},\lambda _4\right),
\nonumber \\
& & v_{\lambda_1\,\lambda_2}^{\alpha \,\beta } (p_1;P)=
u^{\alpha} \left( \frac{P}{2} + {p_1},\lambda_1 \right)
\,v^{\beta } \left(\frac{P}{2} - {p_1},\lambda _2\right) .
\end{eqnarray}
As the independent helicity amplitudes let us choose amplitudes
given in Eq.(6) and proceed to the center mass system.
Opening brackets for the scalar products in Eq.(18), we
obtain a system of 25 linear equations with respect to 25
invariant functions ${f_i}({p_2},{p_1};P)$. This system of
linear equations
was solved by the Gauss approach using a \textsc{Mathematica}
software package. As a result of these manipulations, the
invariant functions can be expressed as linear combinations
of helicity amplitudes:
\begin{eqnarray}
{f_i}(p_2, p_1; P)=
\sum_{\{\lambda_3 \lambda_4, \lambda_1 \lambda_2\}}
 u_{i,\, \{\lambda_3 \lambda_4, \lambda_1 \lambda_2\}}
 \left(p_2,p_1,P\right)
 T_{ \{\lambda_3 \,\lambda _4,\lambda_1\,\lambda_2 \}}
 (p_2,p_1; P) \, ,
\end{eqnarray}
whose transformation coefficients $u_{i,\, \{\lambda_3 \lambda_4, \lambda_1 \lambda_2\}}
\left(p_2,p_1,P\right) $ are given in the
Appendix.
In Eq. (20), the symbol ${\{\lambda_3 \lambda_4, \lambda_1 \lambda_2\}}$
denotes that summation is
not done over all helicity amplitudes but with respect to 25
independent helicity amplitudes only that are given in Eq.(6).
The invariant functions ${g_i}(p_2, p_1; P) $ entering into
Eq.(17) can be
presented in a similar manner:
\begin{eqnarray}
{g_i}(p_2, p_1; P)=
\sum_{\{\lambda_3 \lambda_4, \lambda_1 \lambda_2\}}
 u_{i,\, \{\lambda_3 \lambda_4, \lambda_1 \lambda_2\}}
 \left(p_2,p_1,P\right)
 I_{ \{\lambda_3 \,\lambda _4,\lambda_1\,\lambda_2 \}}(p_2,p_1; P)\, .
\end{eqnarray}
Having derived the matrix for the helicity amplitude
transformation to invariant functions, we may write the
Bethe - Salpeter equation (15) as a system of integral
equations for the invariant functions $ f_{i}$:
\begin{eqnarray}
& & f_{i}(p_2, p_1; P)=g_{i}(p_2, p_1; P)+\sum_{j=1}^{25}
\sum_{\{\lambda_3 \lambda_4, \lambda_1 \lambda_2\} }
\sum_{k=1}^{25} u_{i,\, \{\lambda_3 \lambda_4, \lambda_1 \lambda_2\}}
\left(p_2,p_1,P\right)
{v^*}_{\lambda_3\,\lambda_4}^{\mu \,\nu } (p_2;P)\int
g_{k}(p_2,p';P)\times \nonumber  \\
& & \times R_{\mu \nu \epsilon \eta }^{k} (p_2,p';P)
G^{\epsilon \,\gamma }\,\left( \frac{P}{2} + p' \right)
G^{\eta \,\delta }\,\left( \frac{P}{2} - p' \right)
R_{\gamma \delta \alpha \beta }^{j} (p',p_1;P) f_{j}(p', p_1; P)
\; \frac{d^4 p'}{(2 \pi)^4}\;
v_{\lambda_1\,\lambda_2}^{\alpha \,\beta } (p_1;P) .
\end{eqnarray}
In this equation, the invariant functions $ g_{i}$ and the
vector
particle propagators are calculated using the perturbation
theory based on a Lagrangian interaction, i.e., they may be
obtained to a certain approximation. The system of equations
(22) can be presented in a more concise form:
\begin{eqnarray}
& & f_{i}(p_2, p_1; P)=
g_{i}(p_2, p_1; P)+\sum_{j = 1}^{25} \int K_{i\,j}(p_2,p',p_1;P)
{f_j}(p', p_1; P)\frac{d^4 p'}{(2 \pi)^4}\, ,
\end{eqnarray}
where the kernel $ K_{i\,j}(p_2,p',p_1;P) $ is a 4-scalar 
quantity:
\begin{eqnarray}
& & K_{i\,j}(p_2,p',p_1;P) =\sum_{\{\lambda_3 \lambda_4, \lambda_1 \lambda_2\} }
\sum_{k = 1}^{25} u_{i,\, \{\lambda_3 \lambda_4, \lambda_1 \lambda_2\}}
\left(p_2,p_1,P\right)
{v^*}_{\lambda_3\,\lambda_4}^{\mu \,\nu } (p_2;P)
g_{k}(p_2,p';P)\times \nonumber  \\
& & \times R_{\mu \nu \epsilon \eta }^{k} (p_2,p';P)
G^{\epsilon \,\gamma }\,\left( \frac{P}{2} + p' \right)
G^{\eta \,\delta }\,\left( \frac{P}{2} - p' \right)
R_{\gamma \delta \alpha \beta }^{j} (p',p_1;P)
v_{\lambda_1\,\lambda_2}^{\alpha \,\beta } (p_1;P)\, .
\end{eqnarray}
Thus, the Bethe-Salpeter equation for the vector-particle
scattering amplitude has been reduced to a system of
integral equations for the invariant functions
$f_{i}(p_2, p_1; P)$.

\section{}
\hspace*{\parindent}

In the system of equations (23), integration was carried out
over four independent variables
$ d^4 p' = dp'_0 dp'_1 dp'_2 dp'_3 = dp'_0 p'^2 dp' d\Omega' $.
Using a partial expansion of the invariant function $f_{i}\,$,
we can integrate with
respect to the solid angle $\Omega'$ and reduce the system
of $ 4D $
integral equations , Eq.(23), to a system of $ 2D $ integral
equations for the coefficients of invariant function
expansion with respect to the spherical harmonics.
Let us assume that the $z$-axis of the center mass system is
directed along the initial vector  $\mathbf{p}_1=\frac{1}{2}(\mathbf{k}_1-\mathbf{q}_1) $
and the final vector $\mathbf{p}_2=\frac{1}{2}(\mathbf{k}_2-\mathbf{q}_2) $
corresponds to scattering at an angle $\theta$ in a plane with
a zero azimuthal angle $\phi$. The invariant functions
$f_{i}(p_2, p_1; P)$ depend
on the 4-vectors $p_1, p_2, P$ via the invariants $s$ and $t$
only that are independent of the azimuthal scattering
angle $\phi$.
The expansion of the invariant functions $f_{i}(p_2, p_1; P)$
and $g_{i}(p_2, p_1; P)$
with respect to spherical harmonics will have the form:
\begin{eqnarray}
& & f_{i}(p_2, p_1; P)=
\sum_{l}\frac{1}{|\mathbf{p}_2|}
f_{i}^{l}(p_{2 0},|\mathbf{p}_2|,p_{1 0}, |\mathbf{p}_1|; P)
{Y^{0}_{l}}^{*}(\theta,\,0)\,, \nonumber \\
& & g_{i}(p_2, p_1; P)=
\sum_{l}\frac{1}{|\mathbf{p}_2|} g_{i}^{l}(p_{2 0},|\mathbf{p}_2|,p_{1 0}, |\vec{p_1}|; P)
{Y^{0}_{l}}^{*}(\theta,\,0)\, .
\end{eqnarray}
Although the values $p_{1 0},\,|\mathbf{p}_1|$ and
$p_{2 0},\,|\mathbf{p}_2|$ are not independent for
free particles, they are written in an explicit form in the
arguments of $f_{i}^{l}$ and $g_{i}^{l}$ for the case
where 4-momenta of the
particles are located off the mass surface. The expansion of
the kernel $ K_{i\,j}(p_2,p',p_1;P) $  with respect to
spherical harmonics has the form:
\begin{eqnarray}
K_{i\,j}(p_2,p',p_1;P)=\sum_{l ,\,l' m'}
\frac{1}{|\mathbf{p}_2| |\mathbf{p}'|}
K_{i\,j}^{l 0,\ l' m'}
(p_{2 0},|\mathbf{p}_2|, p_0',|\mathbf{p}'|,p_{1 0},|\mathbf{p}_1|;P)
 {Y^{0}_{l}}^{*}(\theta,\,0)Y^{m'}_{l'}(\theta',\,\phi')\, .
\end{eqnarray}
In this case, as above, we have written the temporal
components of the 4-momenta $p_1$ and $p_2$ as the arguments
of $K_{i\,j}^{l 0,\ l' m'}$, which are independent variables
when the 4-momenta
are not located on the mass surface. Having substituted the
expansions over the spherical harmonics for
$f_{i},\, g_{i}$ and
$K_{i\,j}$ into the system of equations Eq.(23), taking into
account
that the invariant functions ${f_j}(p', p_1; P)$ do not
depend on the
azimuthal angle $\phi'$  of the intermediate
momentum $p'$, we will
arrive at a system of $ 2D $ integral equations for
$f_{i}^{l}$:
\begin{eqnarray}
& & f_{i}^{l}(p_{2 0},|\mathbf{p}_2|,p_{1 0}, |\mathbf{p}_1|; P)=
g_{i}^{l}(p_{2 0},|\mathbf{p}_2|,p_{1 0}, |\mathbf{p}_1|; P)+
\nonumber\\
& & + \sum_{j}\sum_{l'}\int
K_{i\,j}^{l 0,\, l' 0}(p_{2 0},|\mathbf{p}_2|,p_0',|\mathbf{p}'|,
p_{1 0}, |\mathbf{p}_1|;P)
f_{j}^{l'}(p_0',|\mathbf{p}'|;p_{1 0},|\mathbf{p}_1| ;P)
\frac{dp_0'd|\mathbf{p}'|}{(2 \pi)^4}\, .
\end{eqnarray}

\section{}
\hspace*{\parindent}

The functions  $f_{i}^{l}$ entering into Eq.(27) are related
to the
partial reaction amplitudes
$T^{J}_{L',S';\,L,S},\; $ where $J$ is the total momentum,
and $L,\,S$ and $L',\,S'$ are
the orbital momentum and the total spin of the initial and
final states, respectively. The formalism for the helicity
\cite{collins}
is, therefore, more preferable since it makes it
possible to obtain a system of equations for the partial
helicity amplitudes of the reaction
$1^{-}+1^{-}\rightarrow 1^{-}+1^{-}$. The partial helicity
amplitudes of the reaction
$T^{J}_{\lambda_3 \lambda_4, \lambda_1 \lambda_2 }$,
unlike the function $f_{i}^{l}$,
possess a direct physical meaning.

Let us express the spin-1 particle scattering tensor as an
expansion over the helicity amplitudes:
\begin{eqnarray}
& & T_{\mu \nu \alpha \beta }({p_2},{p_1};P)=
\sum_{\lambda_1\,\lambda_2\,\lambda_3 \,\lambda _4}
T_{ \lambda_3 \,\lambda _4, \, \lambda_1\,\lambda_2 }(p_2,p_1; P)
u^{'}_{\mu } \left( \frac{P}{2} + {p_2},\lambda_3 \right)
v^{'}_{\nu } \left(\frac{P}{2} - {p_2},\lambda _4\right)\times
\nonumber\\
& & \times u_{\alpha}^{*} \left( \frac{P}{2} + {p_1},\lambda_1 \right)
\,v_{\beta }^{*} \left(\frac{P}{2} - {p_1},\lambda _2\right)\, ,
\end{eqnarray}
where summation is carried out over 81 helicity amplitudes.
The same representation will be valid for the tensor
$ I_{\mu \nu \alpha \beta }$,
which is the sum of two-body irreducible diagrams:
\begin{eqnarray}
& & I_{\mu \nu \alpha \beta }({p_2},{p_1};P)=
\sum_{\lambda_1\,\lambda_2\,\lambda_3 \,\lambda _4}
I_{ \lambda_3 \,\lambda _4, \, \lambda_1\,\lambda_2 }(p_2,p_1; P)
u^{'}_{\mu } \left( \frac{P}{2} + {p_2},\lambda_3 \right)
v^{'}_{\nu } \left(\frac{P}{2} - {p_2},\lambda _4\right) \times
\nonumber\\
& & \times u_{\alpha}^{*} \left( \frac{P}{2} + {p_1},\lambda_1 \right)
\,v_{\beta }^{*} \left(\frac{P}{2} - {p_1},\lambda _2\right)\, .
\end{eqnarray}

Upon contraction  of the scattering tensor,
Eq.(28), with the
helicity 4-vectors of the initial and final particles, due to
orthogonality of the latter, we will readily obtain the
desired helicity amplitudes. Contrary to the invariant
functions $f_{i}(p_2, p_1; P),\;$, the helicity amplitudes
$T_{ \lambda_3 \,\lambda _4,\,\lambda_1\,\lambda_2 }(p_2,p_1; P)$
depend not only on
the invariants $s$ and $t$ but also
on the azimuthal scattering angle $\phi$.
As shown before, only 25 (Eq.(6)) out of 81 helicity
amplitudes are independent, taking into account the $P-$ and
$T-$ invariance restrictions. Equations (28) and (29) can,
therefore, be written in the form:
\begin{eqnarray}
& & T_{\mu \nu \alpha \beta }({p_2},{p_1};P)=
\sum_{\{ \lambda_1\,\lambda_2\,\lambda_3 \,\lambda _4 \}}
T_{\{ \lambda_3 \,\lambda _4,\, \lambda_1\,\lambda_2 \}}(p_2,p_1; P)
U_{\{ \lambda_3 \,\lambda _4,\,\lambda_1\,\lambda_2 \}
\, \mu \nu \alpha \beta  }({p_2},{p_1};P)\; ,\nonumber\\
& & I_{\mu \nu \alpha \beta }({p_2},{p_1};P)=
\sum_{\{ \lambda_1\,\lambda_2\,\lambda_3 \,\lambda _4 \}}
I_{\{ \lambda_3 \,\lambda _4,\, \lambda_1\,\lambda_2 \}}(p_2,p_1; P)
U_{\{ \lambda_3 \,\lambda _4,\,\lambda_1\,\lambda_2 \}
\, \mu \nu \alpha \beta  }({p_2},{p_1};P)\; ,
\end{eqnarray}
where summation is performed over 25 helicity amplitudes given
in Eq.(6), while the tensors $U_{\{ \lambda_3 \,\lambda _4,\,\lambda_1\,\lambda_2 \}
\, \mu \nu \alpha \beta  }$ are the sum products of the
helicity vector polarizations of the initial and final
particles including the $P-$ and $T-$ invariance relations,
Eqs.(4) and (5). For instance, the tensor
$U_{\{ 1\,1,\, 1\,1 \} \, \mu \nu \alpha \beta  }$
is equal to
\begin{eqnarray}
& & U_{\{ 1\,1,\, 1\,1 \} \, \mu \nu \alpha \beta  }=
u_{\mu}^{'}(1)v_{\nu}^{'}(1) u_{\alpha}^{*}(1)v_{\beta}^{*}(1)+
u_{\mu}^{'}(-1)v_{\nu}^{'}(-1) u_{\alpha}^{*}(-1)v_{\beta}^{*}(-1)\; .
\end{eqnarray}
The other tensors $U_{\{ \lambda_3 \,\lambda _4,\,\lambda_1\,\lambda_2 \}
\, \mu \nu \alpha \beta  }$ in Eq. (30) can be presented in
the same way.
In this formulation, the tensors $T_{\mu \nu  \alpha \beta } $
and $I_{\mu \nu  \alpha \beta } $ will satisfy
the requirements on the $P-$ and $T-$ invariance.

Let us take the following vectors as the helicity 4-vectors
of polarization:
\begin{eqnarray}
& & u^{\mu}(k,1)={\sqrt{\frac{1}{2}}}\,\exp (i \,\phi )\,
( 0,-\cos (\theta )\,\cos (\phi ) + i \,\sin (\phi ),
-i \,\cos (\phi ) - \cos (\theta )\,\sin (\phi ),
\sin (\theta ) )\ , \nonumber  \\
& &u^{\mu}(k,0) =\left( \frac{|\mathbf{k}|}{m},
\frac{k_{0}\,\cos (\phi )\,\sin (\theta )}{m},
\frac{k_{0}\,\sin (\theta )\,\sin (\phi )}{m},
\frac{k_{0}\,\cos (\theta )}{m} \right)\ , \\
& & u^{\mu}(k,-1)={\sqrt{\frac{1}{2}}}\,\exp (-i \,\phi )\,
 ( 0,\cos (\theta )\,\cos (\phi ) + i \,\sin (\phi ),
-i \,\cos (\phi ) + \cos (\theta )\,\sin (\phi ),-\sin
(\theta ) )\; ,
\nonumber
\end{eqnarray}
where $\theta$ and  $\phi $ are the polar and azimuthal
angles of the
vector particle 4-momentum.
For arbitrary 4-momenta, not
necessarily lying on the mass surface, these 4-vectors of
polarization satisfy the relation
\begin{eqnarray}
k \cdot u(k, \lambda)=0, \qquad u(k, \lambda)
\cdot u^{*}(k, \lambda')=
((1+N(k))\delta_{\lambda \, 0}-1) \delta_{\lambda\, \lambda'}\, ,
\end{eqnarray}
where
$\displaystyle{N(k)=\frac{|\mathbf{k}|^{2}-k_{0}^{2}}{m^2}}.$
From these formula it is clear that
4-vectors of polarization, Eq.(32), are orthogonal,
and for $\lambda= \pm 1$ they are normalized with respect
to -1.
It is only in the case of longitudinal polarization
$\lambda = 0\;$
that 4-vectors of polarization are normalized
to $N (k)$ that is equal to -1, if the 4-momentum of the
particle is located on the mass surface.

For this choice of the helicity 4-vectors of polarization,
the scattering tensor $T_{\mu \nu  \alpha \beta }$ satisfies
the relations:
\begin{eqnarray}
& & T_{\mu \nu  \alpha \beta } k_1^{\alpha}=0,\qquad
T_{\mu \nu  \alpha \beta } q_1^{\beta}=0, \nonumber \\
& & T_{\mu \nu  \alpha \beta } k_2^{\mu}=0,\qquad
T_{\mu \nu  \alpha \beta } q_2^{\nu}=0 \; ,
\end{eqnarray}
where $k_1,\;q_1,\;k_2,\;q_2 $ are the 4-momenta of the
initial and final particles.
From Eq. (34) it follows that the fraction
of the vector particle propagator proportional to the product
of its 4-momenta  $k^{\mu} k^{\nu}$ will become zero when
being contracted with the tensors $T_{\mu \nu  \alpha \beta }$
and $I_{\mu \nu  \alpha \beta }$. Hence, upon
convolution with $T_{\mu \nu  \alpha \beta }$ and
$I_{\mu \nu  \alpha \beta }$, the contribution will come
only from a fraction of the propagator that is proportional
to $g^{\mu \nu}$. Note that the Bethe-Salpeter equation now
acquires the following form:
\begin{eqnarray}
& & T_{\mu \nu  \alpha \beta }({p_2},{p_1};P) =
I_{\mu \nu  \alpha \beta }({p_2},{p_1};P) +
\int I_{\mu \nu  \epsilon \eta }({p_2},{p'};P)
T^{\epsilon \eta}_{\phantom{-}\alpha \beta }({p'},{p_1};P) \times
\nonumber \\
& & \phantom{-------} \times
\frac{1}{D(\frac{P}{2}+p')D(\frac{P}{2}-p')}
\frac{d^4 p'}{(2 \pi)^4} \; ,
\end{eqnarray}
where $D(\frac{P}{2}+p')$ and $D(\frac{P}{2}-p')$
are the renormired denominators of
the vector particle
propagators. Using the representations of Eq.(28) and (29)
for the tensors $T_{\mu \nu  \alpha \beta }$ and
$I_{\mu \nu  \alpha \beta }$ and the relations of
orthonormalization  of Eq. (33) for the 4-vectors of
polarization, we may write Eq.(35) as
\begin{eqnarray}
& & T_{\lambda_3 \, \lambda_4, \,\lambda_1 \, \lambda_2 }
({p_2},{p_1};P) = I_{\lambda_3 \, \lambda_4, \,\lambda_1 \,
\lambda_2 }({p_2},{p_1};P) + \sum_{\lambda', \lambda''}
\int I_{\lambda_3 \, \lambda_4, \,\lambda ' \,
\lambda'' }({p_2},{p'};P)
T_{\lambda' \, \lambda'', \,\lambda_1 \,
\lambda_2 }({p'},{p_1};P) \times \nonumber \\
& & \phantom{-------} \times
\frac{((1+N(\frac{P}{2}+p')) \delta_{\lambda' \, 0}-1)
      ((1+N(\frac{P}{2}-p'))\delta_{\lambda'' \, 0}-1) }
{D(\frac{P}{2}+p')D(\frac{P}{2}-p')}\frac{d^4 p'}{(2 \pi)^4} \; .
\end{eqnarray}
In Eq. (36), there is no summation with respect to the
4-indices, instead, the summation is carried out over the
intermediate-particle helicities.  This allows us to move
from the $4D$ system of integral equations (Eq.(36)) to a $2D$
system of integral equations via the expansion of the helicity
amplitudes with respect to
the partial spiral amplitudes and integration over the angles.
The above-mentioned expansion has the following form:
\cite{lifsh}:
\begin{eqnarray}
& & T_{\lambda_3 \, \lambda_4, \,\lambda_1 \,\lambda_2 }({p_2},{p_1};P)=
\sum_{J M}\frac{2 J + 1}{4\pi}D^{J}_{\lambda_3-\lambda_4,\; M}
(\mathbf{n}_2)D^{J^{\scriptstyle * }}_{\lambda_1-\lambda_2,\; M}
(\mathbf{n}_1) \times \nonumber \\
& & \phantom{--------} \times T^{J}_{\lambda_3 \, \lambda_4,
\,\lambda_1 \,\lambda_2 }
(p_{20},|\mathbf{p}_2|,p_{10},|\mathbf{p}_1|; P) \; .
\end{eqnarray}

\begin{eqnarray}
& & I_{\lambda_3 \, \lambda_4, \,\lambda_1 \,\lambda_2 }({p_2},{p_1};P)=
\sum_{J M}\frac{2 J + 1}{4\pi}D^{J}_{\lambda_3-\lambda_4,\; M}
(\mathbf{n}_2)D^{J^{\scriptstyle * }}_{\lambda_1-\lambda_2,\; M}
(\mathbf{n}_1) \times \nonumber \\
& & \phantom{------} \times I^{J}_{\lambda_3 \, \lambda_4, \,\lambda_1 \,\lambda_2 }
(p_{20},|\mathbf{p}_2|,p_{10},|\mathbf{p}_1|; P) \; .
\end{eqnarray}

In these expressions, the $D-$ function arguments are taken 
as
\begin{equation}
D^{J}_{\Lambda \; M}(\mathbf{n}) =
D^{J}_{\Lambda \; M}(\phi, \theta, 0)=
e^{i M \phi} d^{J}_{\Lambda \; M}(\theta)\; ,
\end{equation}
where $\theta$ and  $\phi$ are the polar and azimuthal angles
of the unit vector $\mathbf{n}$. Using the orthogonality
properties of the $D-$ functions, we may integrate with
respect to the angular variables in Eq.(36). As a result, the
system of $4D$ integral equations for the helicity amplitudes,
Eq.(36),  will be reduced to a system of $2D$ integral equations
for the partial helicity amplitudes:
\begin{eqnarray}
& & T^{J}_{\lambda_3 \, \lambda_4, \,\lambda_1 \,\lambda_2 }
(p_{20},|\mathbf{p}_{2}|,p_{10},|\mathbf{p}_1|; P)=
I^{J}_{\lambda_3 \, \lambda_4, \,\lambda_1 \,\lambda_2 }
(p_{20},|\mathbf{p}_{2}|,p_{10},|\mathbf{p}_1|; P)+ \nonumber \\
& & + \sum_{\lambda' \lambda''} \int
I^{J}_{\lambda_3 \, \lambda_4, \,\lambda' \,\lambda'' }
(p_{20},|\mathbf{p}_2|,p'_{0},|\mathbf{p}'|; P)
T^{J}_{\lambda' \, \lambda'', \,\lambda_1 \,\lambda_2 }
(p'_{0},|\mathbf{p}'|,p_{10},|\mathbf{p}_1|; P) \times \nonumber \\
& &\times \frac{ ((N(\frac{P}{2}+p'))\delta_{\lambda' \, 0}-1)
     ((N(\frac{P}{2}-p'))\delta_{\lambda'' \, 0}-1)}
{D(\frac{P}{2}+p')D(\frac{P}{2}-p')}
\frac{|\mathbf{p}'|^2 \, dp'_0 \, d|\mathbf{p}'|}{(2 \pi)^4} \; .
\end{eqnarray}
The partial spiral amplitudes satisfy the relations for
the $T-$ and $P-$ invariance:
\begin{eqnarray}
& & T^{J}_{\lambda_3 \, \lambda_4, \,\lambda_1 \,\lambda_2 }=
\eta_1 \eta_2 \eta_3 \eta_4 (-1)^{s_3+s_4-s_1-s_2}
T^{J}_{-\lambda_3 \, -\lambda_4, \,-\lambda_1 \,-\lambda_2 },\nonumber \\
& & T^{J}_{\lambda_3 \, \lambda_4, \,\lambda_1 \,\lambda_2 }=
T^{J}_{\lambda_1 \, \lambda_2, \,\lambda_3 \,\lambda_4 }.
\end{eqnarray}
This is the reason why only 25 out of 81 partial helicity
amplitudes with this $J$ will be independent. Let us select
the partial helicity amplitudes with the same helicity as in
Eq.(6) as 25 independent amplitudes and number them from 1
to 25. Now, the system of $2D$ integral equations can be
written in terms of 25 independent partial amplitudes as
follows:
\begin{eqnarray}
& & \phantom{------} T^{J}_{i}(p_{20},|\mathbf{p}_{2}|,p_{10},|\mathbf{p}_1|; P)=
I^{J}_{i}(p_{20},|\mathbf{p}_{2}|,p_{10},|\mathbf{p}_1|; P) +  \\
& & + \sum_{j=1}^{25} \int
K^{J}_{i, j}(p_{20},|\mathbf{p}_{2}|,p'_{0},|\mathbf{p}'|; P)
T^{J}_{j}(p'_{0},|\mathbf{p}'|,p_{10},|\mathbf{p}_1|; P)
\frac{ |\mathbf{p}'|^2 }
{D(\frac{P}{2}+p')D(\frac{P}{2}-p')}
\frac{ dp'_0 \, d|\mathbf{p'}|}{(2 \pi)^4}. \nonumber
\end{eqnarray}
The values $ K^{J}_{i, j} $ are expressed via
$ I^{J}_{i, j} $
and the normalization multipliers $N$. Their explicit
representation is given in Appendix.

This work was supported by the Russian Basic Research
Foundation (grants N N \,  01-02-17276 and 02-02-06866) and
the RF Ministry of Education (grant N \, E00-3, 3-110).

\vspace{3 cm}

\appendix{\LARGE{\textsc{Appendix}}}
\vspace{1 cm}\\
In the general case, the matrix for the transformation of 25
helicity amplitudes to 25 invariant functions has 625 elements.
Some of these elements are considered to be cumbersome
mathematical expressions. In this appendix we present the
transformation matrix elements for scattering of the vector
particles having equal masses $ m_1 = m_2 = m $.  In this case,
out of 625 elements, 361 are equal to zero, and the remaining
264 non-zero elements are related by numerous symmetry
relations.
The non-zero elements may be classified into 63 groups with
their elements differing in the constant factors only.
For the sake of space, the spiral amplitude indices
corresponding to Eq.(6) were numbered from 1 to 25.
Given below are the explicit expressions for the non-zero
transformation matrix elements $u_{i\; j},\;$  where $i - $
is the index of
the invariant function $f_{i}, \;$ and $j - $ is the helicity
amplitude index.
\begin{eqnarray}
& & u_{1\, 1}=\frac{1}{2}u_{1\, 3}=\frac{1}{2} u_{1\, 7}=u_{1\, 9}=-u_{2\, 3}=
u_{2\, 7}=-u_{3\, 3}=u_{3\, 7}=2u_{4\, 3}=2 u_{4 \,7}= 2u_{6\, 3}=
\nonumber   \\
& & = 2u_{6\, 7}=\frac{-8\,{\csc (\theta )}^2}{{\left( -4\,m^2 + s \right) }^2} \  , \nonumber  \\
& & u_{4\, 1}=u_{4\, 9}=\frac{{\csc (\frac{\theta }{2})}^2\,{\sec (\frac{\theta }{2})}^2}
{-8\,m^2\,s + 2\,s^2}
\ ,\nonumber   \\
& & u_{5\, 1}=u_{8\, 1}=\frac{-\left( \left( -4\,m^2 + s +
        \left( 4\,m^2 - 2\,s \right) \,\cos (\theta ) \right) \,
      {\csc (\frac{\theta }{2})}^2\,{\sec (\frac{\theta }{2})}^2 \right) }
{2\,s\,{\left( -4\,m^2 + s \right) }^2}
\ , \nonumber  \\
& & u_{6 \,1}=\frac{\left( -4\,m^2 + s\,\cos (\theta ) \right) \,
    {\csc (\frac{\theta }{2})}^2\,{\sec (\frac{\theta }{2})}^2}{2\,s\,
    {\left( -4\,m^2 + s \right) }^2}
\ , \nonumber  \\
& & u_{7\, 1}=-u_{9\, 1}=\frac{\left( -1 + 3\,\cos (\theta ) \right) \,{\csc (\frac{\theta }{2})}^2\,
    {\sec (\frac{\theta }{2})}^2}{4\,\left( 4\,m^2\,s - s^2 \right) }
    \ ,\nonumber \\
& & u_{10\, 1}=\frac{-\left( \left( -8\,m^4 - 8\,m^2\,s + 3\,s^2
+ 2\,\left( 4\,m^2 - s \right) \,s\,\cos (\theta ) +
4\,\left( 2\,m^4 - 4\,m^2\,s + s^2 \right) \,\cos (2\,\theta ) \
\right)
\right) }{8\,s^2\,{\left( -4\,m^2 + s \right) }^2}
\times \nonumber  \\
& & \times {\csc (\frac{\theta }{2})}^2\,
{\sec (\frac{\theta }{2})}^2\ ,\nonumber  \\
& &  u_{11\, 1}=u_{11\,9}=-u_{11\, 17}=-u_{11\, 21}=u_{18\, 1}=
-u_{18\, 9}=-u_{18\, 17}=u_{18\, 21}=\frac{{\csc (\frac{\theta }{2})}^2\,{\sec (\frac{\theta }{2})}^2}
  {8\,m^2 - 2\,s}\ , \nonumber  \\
& & u_{12\,1}=u_{12\,9}=-u_{12\,17}=-u_{12\,21}=-u_{13\,1}=
-u_{13\,9}=u_{13\,17}=u_{13\,21}=u_{19\,1}=u_{19\,9}=-u_{19\,17}=
\nonumber  \\
& & = u_{19\,21}= -u_{20\,1}=u_{20\,9}=u_{20\,17}=-u_{20\,21}=\frac{\cos (\theta )\,{\csc (\frac{\theta }{2})}^2\,
    {\sec (\frac{\theta }{2})}^2}{-16\,m^2 + 4\,s}\ ,\nonumber  \\
& & u_{14\,1}=u_{14\,9}=-u_{14\,17}=-u_{14\,21}=u_{21\,1}=-u_{21\,9}=
-u_{21\,17}=u_{21\,21}=\frac{{\cos (\theta )}^2\,{\csc (\frac{\theta }{2})}^2\,
    {\sec (\frac{\theta }{2})}^2}{-32\,m^2 + 8\,s}\ , \nonumber  \\
& & u_{15 \,1}=\frac{1}{2}u_{15 \,3}=u_{15 \,17}=-2u_{16\,1}=
-u_{16 \,3}=-2u_{16 \,17}=u_{22 \,1}=\frac{1}{2}u_{22 \,7}=
u_{22\,17}=2u_{23\,1}=u_{23 \,7}=\nonumber  \\
& & = 2 u_{23\,17}=\frac{{\sec (\frac{\theta }{2})}^2}{-4\,m^2 + s}\ , \nonumber  \\
& & u_{17\,1}=\frac{1}{2}u_{17\,3}=u_{17\,17}=u_{24\,1}=
\frac{1}{2}u_{24\,7}=u_{24\,17}=\frac{-\left( \left( -4\,m^2 + s +
\left( -4\,m^2 + 2\,s \right) \,\cos (\theta ) \right) \,
{\sec (\frac{\theta }{2})}^2 \right) }{4\,s\,
\left( -4\,m^2 + s \right) }\ , \nonumber  \\
& & u_{25\,1}=\frac{1}{2}u_{25\,3}=\frac{1}{2}u_{25\,7}=
u_{25\,9}=u_{25\,17}=u_{25\,21}=\frac{1}{2}\ , \nonumber  \\
& & u_{5\,2}=u_{5\,8}=2u_{6\,2}=2u_{6\,4}=2u_{6\,6}=2u_{6\,8}=
4u_{7\,8}=u_{8\,4}=u_{8\,6}=-4u_{9\,6}=\frac{8\,{\sqrt{2}}\,m\,\csc (\theta )}
{{\sqrt{s}}\,{\left( -4\,m^2 + s \right) }^2}\ , \nonumber  \\
& & u_{7\,2}=-u_{9\,4}=\frac{2\,{\sqrt{2}}\,m\,\left( 8\,m^2 - 3\,s \right) \,\csc (\theta )}
{s^{\frac{3}{2}}\,{\left( -4\,m^2 + s \right) }^2}\ , \nonumber  \\
& & u_{7\,4}=u_{7\,6}=-u_{9\,2}=-u_{9\,8}=\frac{-2\,{\sqrt{2}}\,m\,\csc (\theta )}
{s^{\frac{3}{2}}\,\left( -4\,m^2 + s \right) }\ , \nonumber  \\
& & u_{10\,2}=u_{10\,4}=\frac{8\,{\sqrt{2}}\,m\,\left( 3\,m^2 - s \right) \,\cot (\theta )}
{s^{\frac{3}{2}}\,{\left( -4\,m^2 + s \right) }^2}\ , \nonumber  \\
& & u_{12\,2}=-u_{12\,8}=-u_{12\,11}=-u_{12\,15}=-u_{13\,4}=
u_{13\,6}=-u_{13\,18}=u_{13\,20}=u_{16\,4}=u_{16\,6}=
\nonumber  \\ & & = u_{16\,18}=
u_{16\,20}= u_{19\,2}=u_{19\,8}=-u_{19\,11}=u_{19\,15}=-u_{20\,4}=
-u_{20\,6}=u_{20\,18}=u_{20\,20}=-u_{23\,2}=\nonumber  \\ & &
 = -u_{23\,8}=-u_{23\,11}=u_{23\,15}=\frac{2\,{\sqrt{2}}\,m\,\csc (\theta )}
{{\sqrt{s}}\,\left( -4\,m^2 + s \right) }\ , \nonumber  \\
& & u_{14\,2}=u_{14\,4}=-u_{14\,6}=-u_{14\,8}=-u_{14\,11}=-u_{14\,15}=
u_{14\,18}=-u_{14\,20}=\frac{1}{2}u_{17\,4}=\frac{1}{2}u_{17\,6}=
\frac{1}{2}u_{17\,18}= \nonumber  \\
& & = \frac{1}{2}u_{17\,20}=u_{21\,2}=u_{21\,4}=
u_{21\,6}=u_{21\,8}=-u_{21\,11}=u_{21\,15}=-u_{21\,18}=-u_{21\,20}=
\frac{1}{2}u_{24\,2}=\frac{1}{2}u_{24\,8}= \nonumber  \\
& & = \frac{1}{2}u_{24\,11}=
-\frac{1}{2}u_{24\,15}=\frac{{\sqrt{2}}\,m\,\cot (\theta )}{{\sqrt{s}}\,\left( -4\,m^2 + s \right) }
\ , \nonumber  \\
& & u_{5\,3}=-2 u_{7\,3}=u_{8\,7}=2 u_{9\,7}=
\frac{{\sqrt{2}}\,m\,\cot (\theta )}{{\sqrt{s}}\,\left( -4\,m^2 + s \right) }\ , \nonumber  \\
& & u_{5\,7}=2 u_{7\,7}=u_{8\,3}=-2 u_{9\,3}=\frac{\left( -4\,m^2 + s + \left( -4\,m^2 + 2\,s \right) \,
\cos (\theta ) \right) \,{\csc (\frac{\theta }{2})}^2\,
{\sec (\frac{\theta }{2})}^2}{s\,{\left( -4\,m^2 + s \right) }^2}\ , \nonumber  \\
& & u_{10\,3}=u_{10\,7}=\frac{\left( -4\,m^2 + s + 4\,m^2\,\cos (\theta ) \right) \,
\left( -4\,m^2 + s + \left( -4\,m^2 + 2\,s \right) \,
\cos (\theta ) \right) \,{\csc (\frac{\theta }{2})}^2\,
{\sec (\frac{\theta }{2})}^2}{4\,s^2\,{\left( -4\,m^2 + s \right) }^2}\ , \nonumber  \\
& & u_{15\,7}=2u_{15\,7}=2u_{15\,21}=2 u_{16\,7}=4 u_{16\,9}=
4 u_{16\,21}=u_{22\,3}=2 u_{22\,9}=2 u_{22\,21}=-2 u_{23\,3}=
-4u_{23\,9}= \nonumber  \\
& & = -4 u_{23\,21}=\frac{-2\,{\csc (\frac{\theta }{2})}^2}{-4\,m^2 + s}
\ , \nonumber  \\
& & u_{17\,7}=2u_{17\,9}=2u_{17\,21}=u_{24\,3}=2u_{24\,9}=2u_{24\,21}=
\frac{\left( -4\,m^2 + s + 4\,m^2\,\cos (\theta ) \right) \,
    {\csc (\frac{\theta }{2})}^2}{2\,\left( 4\,m^2\,s - s^2 \right) } \ , \nonumber  \\
& & u_{10\,5}=\frac{32\,m^4 - 16\,m^2\,s}{{\left( -4\,m^2\,s + s^2 \right) }^2} \ , \nonumber  \\
& & u_{14\,5}=u_{14\,19}=-u_{17\,22}=-u_{17\,24}=-u_{21\,12}=
-u_{21\,14}=-u_{24\,10}=-u_{24\,16}=\frac{4\,m^2}{-4\,m^2\,s + s^2} \ , \nonumber  \\
& & u_{10\,6}=u_{10\,8}=-\frac{1}{2}u_{10\,13}=-\frac{1}{2}u_{10\,23}=
\frac{8\,{\sqrt{2}}\,m^3\,\cot (\theta )}
  {s^{\frac{3}{2}}\,{\left( -4\,m^2 + s \right) }^2}\ , \nonumber  \\
& & u_{5\,10}=u_{6\,14}=-2u_{7\,10}=2u_{7\,14}=u_{8\,22}=-2u_{9\,14}=
2u_{9\,22}=\frac{-8\,m^2\,{\sec (\frac{\theta }{2})}^2}
  {s\,{\left( -4\,m^2 + s \right) }^2}\ , \nonumber  \\
& & u_{10\,9}=\frac{\left( 8\,m^4 - 8\,m^2\,s + s^2 -
      2\,\left( 4\,m^2 - s \right) \,s\,\cos (\theta ) -
      8\,m^4\,\cos (2\,\theta ) \right) \,{\csc (\frac{\theta }{2})}^2\,
    {\sec (\frac{\theta }{2})}^2}{8\,s^2\,{\left( -4\,m^2 + s \right) }^2}\ , \nonumber  \\
& & u_{10\,10}= u_{10\,22}=\frac{-2\,m^2\,\left( 4\,m^2 - s +
      \left( 4\,m^2 - 2\,s \right) \,\cos (\theta ) \right) \,
    {\sec (\frac{\theta }{2})}^2}{s^2\,{\left( -4\,m^2 + s \right) }^2}
\ , \nonumber  \\
& & u_{5\,9}=u_{6\,9}=u_{7\,9}=u_{8\,9}=u_{9\,9}=
\frac{\left( -4\,m^2 + s - 4\,m^2\,\cos (\theta ) \right) \,
    {\csc (\frac{\theta }{2})}^2\,{\sec (\frac{\theta }{2})}^2}{2\,s\,
    {\left( -4\,m^2 + s \right) }^2}\ , \nonumber  \\
& & u_{2\,11}=-u_{3\,20}=-2u_{4\,11}=-2u_{4\,20}=\frac{16\,{\sqrt{2}}\,m\,\left( \cot (\theta ) - \csc (\theta ) \right) \,
    {\csc (\theta )}^2}{{\sqrt{s}}\,{\left( -4\,m^2 + s \right) }^2}
\ , \nonumber  \\
& & u_{5\,11}=-2u_{6\,11}=-2u_{6\,20}=u_{8\,20}=\frac{-4\,{\sqrt{2}}\,m\,{\sec (\frac{\theta }{2})}^2\,
    \tan (\frac{\theta }{2})}{{\sqrt{s}}\,{\left( -4\,m^2 + s \right) }^2}
\ , \nonumber  \\
& & u_{7\,11}=-u_{9\,20}=- \frac{m\,\left( 8\,m^2 - 3\,s +
        \left( 8\,m^2 + s \right) \,\cos (\theta ) \right) \,
      \csc (\frac{\theta }{2})\,{\sec (\frac{\theta }{2})}^3}{{\sqrt{2}}\,
      s^{\frac{3}{2}}\,{\left( -4\,m^2 + s \right) }^2}
\ , \nonumber  \\
& & u_{7\,20}=-u_{9\,11}=\frac{m\,\left( 4\,m^2 - s + \left( 4\,m^2 - 3\,s \right) \,
       \cos (\theta ) \right) \,\csc (\frac{\theta }{2})\,
    {\sec (\frac{\theta }{2})}^3}{{\sqrt{2}}\,s^{\frac{3}{2}}\,
    {\left( -4\,m^2 + s \right) }^2}
\ , \nonumber  \\
& & u_{10\,11}=u_{10\,20}=-\frac{{\sqrt{2}}\,m\,\cos (\theta )\,
      \left( 2\,m^2 - s + 2\,m^2\,\cos (\theta ) \right) \,
      \csc (\frac{\theta }{2})\,{\sec (\frac{\theta }{2})}^3}{
      s^{\frac{3}{2}}\,{\left( -4\,m^2 + s \right) }^2}
\ , \nonumber  \\
& &  u_{5\,16}=u_{6\,12}=-2u_{7\,12}=2u_{7\,16}=u_{8\,24}=
2u_{9\,12}=-2u_{9\,24}=\frac{8\,m^2\,{\csc (\frac{\theta }{2})}^2}
  {s\,{\left( -4\,m^2 + s \right) }^2}\ , \nonumber  \\
& & u_{10\,12}=\frac{4\,m^2\,\left( -4\,m^2 + s +
\left( 4\,m^2 - 2\,s \right) \,\cos (\theta ) \right) \,
{\csc (\frac{\theta }{2})}^2}{{\left( -4\,m^2\,s + s^2 \right) }^2}
 \ , \nonumber  \\
& &  u_{2\,15}=u_{3\,18}=2u_{4\,15}=-2u_{4\,18}=\frac{-16\,{\sqrt{2}}\,m\,{\csc (\theta )}^2\,
    \left( \cot (\theta ) + \csc (\theta ) \right) }{{\sqrt{s}}\,
    {\left( -4\,m^2 + s \right) }^2}\ , \nonumber  \\
& &  u_{5\,15}=-2u_{6\,15}=2u_{6\,18}=-u_{8\,18}=\frac{4\,{\sqrt{2}}\,m\,\cot (\frac{\theta }{2})\,
    {\csc (\frac{\theta }{2})}^2}{{\sqrt{s}}\,
    {\left( -4\,m^2 + s \right) }^2} \ , \nonumber  \\
& &  u_{7\,15}=u_{9\,18}=\frac{m\,\left( 1 + 3\,\cos (\theta ) \right) \,
    {\csc (\frac{\theta }{2})}^3\,\sec (\frac{\theta }{2})}{{\sqrt{2}}\,
    {\sqrt{s}}\,{\left( -4\,m^2 + s \right) }^2}
\ , \nonumber  \\
& &  u_{7\,18}=u_{9\,15}=-\frac{m\,\left( -4\,m^2 + s +
        \left( 4\,m^2 + s \right) \,\cos (\theta ) \right) \,
      {\csc (\frac{\theta }{2})}^3\,\sec (\frac{\theta }{2})}{{\sqrt{2}}\,
      s^{\frac{3}{2}}\,{\left( -4\,m^2 + s \right) }^2}
\ , \nonumber  \\
& &  u_{7\,13}=-u_{9\,13}=\frac{-16\,{\sqrt{2}}\,m^3\,\csc (\theta )}
  {s^{\frac{3}{2}}\,{\left( -4\,m^2 + s \right) }^2}
\ , \nonumber  \\
& &  u_{10\,14}=\frac{-4\,m^2\,\left( 4\,m^2 - s + 4\,m^2\,\cos (\theta ) \right) \,
    {\sec (\frac{\theta }{2})}^2}{s^2\,{\left( -4\,m^2 + s \right) }^2}
\ , \nonumber  \\
& &  u_{10\,15}=-u_{10\,18}=\frac{{\sqrt{2}}\,m\,\cos (\theta )\,
\left( -2\,m^2 + s +
2\,m^2\,\cos (\theta ) \right) \,{\csc (\frac{\theta }{2})}^3\,
\sec (\frac{\theta }{2})}{s^{\frac{3}{2}}\,
{\left( -4\,m^2 + s \right) }^2}\ , \nonumber  \\
& &  u_{10\,16}=u_{10\,24}=\frac{2\,m^2\,\left( -4\,m^2 + s + 4\,m^2\,\cos (\theta ) \right) \,
    {\csc (\frac{\theta }{2})}^2}{{\left( -4\,m^2\,s + s^2 \right) }^2}
\ , \nonumber  \\
& &  u_{1\,17}=-\frac{\left( -3 + \cos (\theta ) \right) \,
      {\csc (\frac{\theta }{2})}^2\,{\sec (\frac{\theta }{2})}^4}
      {{\left( -4\,m^2 + s \right) }^2}
\ , \nonumber  \\
& &  u_{2\,17}=-u_{3\,17}=\frac{2\,{\sec (\frac{\theta }{2})}^4}{{\left( -4\,m^2 + s \right) }^2}
\ , \nonumber  \\
& &  u_{4\,17}=\frac{\left( 4\,m^2 - 3\,s +
      \left( 4\,m^2 + s \right) \,\cos (\theta ) \right) \,
    {\csc (\frac{\theta }{2})}^2\,{\sec (\frac{\theta }{2})}^4}{4\,s\,
    {\left( -4\,m^2 + s \right) }^2}
\ , \nonumber  \\
& &  u_{5\,17}=u_{8\,17}=\frac{\left( 2\,m^2 - 3\,s\,\cos (\theta ) +
      \left( -2\,m^2 + s \right) \,\cos (2\,\theta ) \right) \,
    {\csc (\frac{\theta }{2})}^2\,{\sec (\frac{\theta }{2})}^4}{4\,s\,
    {\left( -4\,m^2 + s \right) }^2}
\ , \nonumber  \\
& &  u_{6\,17}=\frac{\left( 8\,m^2 - 5\,s + 2\,\left( 4\,m^2 + s \right) \,
       \cos (\theta ) - s\,\cos (2\,\theta ) \right) \,
    {\csc (\frac{\theta }{2})}^2\,{\sec (\frac{\theta }{2})}^4}{8\,s\,
    {\left( -4\,m^2 + s \right) }^2}\ , \nonumber  \\
& & u_{7\,17}=-u_{9\,17}=\frac{\left( -12\,m^2 + s + 8\,\left( -2\,m^2 + s \right) \,
       \cos (\theta ) - \left( 4\,m^2 + s \right) \,\cos (2\,\theta ) \
\right) \,{\csc (\frac{\theta }{2})}^2\,{\sec (\frac{\theta }{2})}^4}{16\,
    s\,{\left( -4\,m^2 + s \right) }^2}
\ , \nonumber  \\
& & u_{10\,17}=\frac{\left( \left( 4\,m^4 - 16\,m^2\,s + 3\,s^2 \right) \,
\cos (\theta ) + \left( -8\,m^4 - 4\,m^2\,s + 3\,s^2 \right) \,
\cos (2\,\theta )\right)
\csc (\frac{\theta }{2})^2\,\sec (\frac{\theta }{2})^4 }
{16\,s^2\,{\left( -4\,m^2 + s \right) }^2}+\nonumber  \\
& & + \frac {\left( 4\,
\left( 2\,m^4 - 3\,m^2\,s + s^2 - m^4\,\cos (3\,\theta ) \right)  \
\right) \,{\csc (\frac{\theta }{2})}^2\,{\sec (\frac{\theta }{2})}^4}
{16\,s^2\,{\left( -4\,m^2 + s \right) }^2}
\ , \nonumber  \\
& & u_{4\,19}=\frac{32\,m^2\,{\csc (\theta )}^2}{s\,{\left( -4\,m^2 + s \right) }^2}
\ , \nonumber  \\
& & u_{7\,19}=-u_{9\,19}=\frac{-16\,m^2\,\cot (\theta )\,\csc (\theta )}
  {s\,{\left( -4\,m^2 + s \right) }^2}
\ , \nonumber  \\
& & u_{10\,19}=\frac{-16\,m^2\,\left( -m^2 + s + m^2\,\cos (2\,\theta ) \right) \,
    {\csc (\theta )}^2}{s^2\,{\left( -4\,m^2 + s \right) }^2}
\ , \nonumber  \\
& & u_{1\,21}=\frac{\left( 3 + \cos (\theta ) \right) \,{\csc (\frac{\theta }{2})}^4\,
    {\sec (\frac{\theta }{2})}^2}{{\left( -4\,m^2 + s \right) }^2}
\ , \nonumber  \\
& & u_{2\,21}=-u_{3\,21}=\frac{-2\,{\csc (\frac{\theta }{2})}^4}{{\left( -4\,m^2 + s \right) }^2}
\ , \nonumber  \\
& & u_{4\,21}=\frac{-\left( \left( -4\,m^2 + 3\,s +
        \left( 4\,m^2 + s \right) \,\cos (\theta ) \right) \,
      {\csc (\frac{\theta }{2})}^4\,{\sec (\frac{\theta }{2})}^2 \right) }\
{4\,s\,{\left( -4\,m^2 + s \right) }^2}\ , \nonumber  \\
& & u_{5\,21}=u_{8\,21}=\frac{\left( -2\,m^2 + s + 3\,s\,\cos (\theta ) +
      2\,m^2\,\cos (2\,\theta ) \right) \,{\csc (\frac{\theta }{2})}^4\,
    {\sec (\frac{\theta }{2})}^2}{4\,s\,{\left( -4\,m^2 + s \right) }^2}
\ , \nonumber  \\
& & u_{6\,21}=\frac{-\left( \left( 8\,m^2 + s +
        \left( -8\,m^2 + 6\,s \right) \,\cos (\theta ) +
        s\,\cos (2\,\theta ) \right) \,{\csc (\frac{\theta }{2})}^4\,
      {\sec (\frac{\theta }{2})}^2 \right) }{8\,s\,
    {\left( -4\,m^2 + s \right) }^2}
 \ , \nonumber  \\
& & u_{7\,21}=-u_{9\,21}=\frac{\left( -4\,m^2 + s + \left( 4\,m^2 + s \right) \,\cos (\theta ) \
\right) \,{\csc (\frac{\theta }{2})}^4}{4\,s\,
    {\left( -4\,m^2 + s \right) }^2}
 \ , \nonumber  \\
& & u_{10\,21}=\left( \left( -4\,m^4 + s^2 \right) \,\cos (\theta ) +
\left( -8\,m^4 + 4\,m^2\,s + s^2 \right) \,\cos (2\,\theta ) +
\right.\nonumber\\ & & \left. + 2\,\left( 4\,m^4 - 2\,m^2\,s + s^2 + 2\,m^4\,\cos (3\,\theta ) \
\right)  \right)\frac{\csc (\frac{\theta }{2})^4\,
\sec (\frac{\theta }{2})^2}
{16\,s^2\,{\left( -4\,m^2 + s \right) }^2}
\nonumber  \\
& & u_{10\,25}=\frac{16\,m^4}
{{\left( -4\,m^2\,s + s^2 \right) }^2}\nonumber
\end{eqnarray}
In these formulas $\theta $ is the scattering angle and $s$ is the
squared total energy in the center of mass system.

Out of 625 elements of the matrix $ K^{J}_{i, j},\, $ 420
appear to be equal to zero. Presented below are 205 non-zero 
elements of the $ K^{J}_{i, j} $. The order of numbering
corresponds to the order in which the helicity amplitudes are
given in Eq. (6).
\begin{eqnarray}
& & K_{1,1}^{J}=I_{1}^{J}, \; K_{1,2}^{J}=N_2 I_{2}^{J}, \;
K_{1,3}^{J}=I_{3}^{J},\; K_{1,4}^{J}=N_1 I_{4}^{J}, \;
K_{1,5}^{J}=N_1 N_2 I_{5}^{J},\;K_{1,6}^{J}=N_1 I_{6}^{J},\;
K_{1,7}^{J}=I_{7}^{J}, \; \nonumber  \\
& & K_{1,8}^{J}=N_2 I_{8}^{J},\;
K_{1,9}^{J}=I_{9}^{J}; \nonumber  \\
& & K_{2,2}^{J}=I_{1}^{J}, \; K_{2,8}^{J}=-I_{9}^{J}, \;
K_{2,10}^{J}=-N_2 I_{2}^{J}, \; K_{2,11}^{J}=-I_{3}^{J}, \;
K_{2,12}^{J}=-N_1 I_{4}^{J}, \;
K_{2,13}^{J}=-N_1 N_2 I_{5}^{J},\; \nonumber  \\
& & K_{2,14}^{J}=-N_1 I_{6}^{J},\; K_{2,15}^{J}=-I_{7}^{J}, \;
K_{2,16}^{J}=-N_2 I_{8}^{J};  \nonumber  \\
& & K_{3,3}^{J}=I_{1}^{J}, \;K_{3,7}^{J}=I_{9}^{J}, \;
K_{3,11}^{J}=-N_2 I_{2}^{J}, \;K_{3,15}^{J}=N_2 I_{8}^{J}, \;
K_{3,17}^{J}=I_{3}^{J}, \;K_{3,18}^{J}=N_1 I_{4}^{J}, \;
\nonumber  \\ & &
K_{3,19}^{J}=N_1 N_2 I_{5}^{J}, \;
K_{3,20}^{J}=N_1 I_{6}^{J}, \; K_{3,21}^{J}=I_{7}^{J};\nonumber  \\
& & K_{4,4}^{J}=I_{1}^{J}, \; K_{4,6}^{J}=-I_{9}^{J}, \;
K_{4,12}^{J}=-N_2 I_{2}^{J}, \;K_{4,14}^{J}=-N_2 I_{8}^{J},\;
K_{4,18}^{J}=I_{3}^{J}, \;K_{4,20}^{J}=-I_{7}^{J},
\nonumber  \\ & &
K_{4,22}^{J}=-N_1 I_{4}^{J},\;
K_{4,23}^{J}=-N_1 N_2 I_{5}^{J},\; K_{4,24}^{J}=-N_1 I_{6}^{J};\nonumber  \\
& & K_{5,5}^{J}=I_{1}^{J}+I_{9}^{J}, \;
K_{5,13}^{J}=N_2 ( I_{8}^{J}-I_{2}^{J}),\;
K_{5,19}^{J}=I_{3}^{J}+I_{7}^{J},\;
K_{5,23}^{J}=N_1 ( I_{6}^{J}-I_{4}^{J}),\;
K_{5,25}^{J}=N_1 N_2 I_{5}^{J};\nonumber  \\
& & K_{6,4}^{J}=-I_{9}^{J}, \;K_{6,6}^{J}=I_{1}^{J}, \;
K_{6,12}^{J}=-N_2 I_{8}^{J}, \;K_{6,14}^{J}=-N_2 I_{2}^{J}, \;
K_{6,18}^{J}=-I_{7}^{J}, \;K_{6,20}^{J}=I_{3}^{J}, \;
\nonumber  \\ & &
K_{6,22}^{J}=-N_1 I_{6}^{J},
K_{6,23}^{J}=N_1 N_2 I_{5}^{J}, \;
K_{6,24}^{J}=-N_1 I_{4}^{J};\nonumber  \\
& & K_{7,3}^{J}=I_{9}^{J},\; K_{7,7}^{J}=I_{1}^{J},\;
K_{7,11}^{J}=N_2 I_{8}^{J},\;K_{7,15}^{J}=-N_2 I_{2}^{J},\;
K_{7,17}^{J}=I_{7}^{J},\; K_{7,18}^{J}=-N_1 I_{6}^{J},\;
\nonumber  \\ & &
K_{7,19}^{J}=N_1 N_2 I_{5}^{J},
K_{7,20}^{J}=-N_1 I_{4}^{J};\nonumber  \\
& & K_{8,2}^{J}=-I_{9}^{J},\; K_{8,8}^{J}=I_{1}^{J},\;
K_{8,10}^{J}=-N_2 I_{8}^{J},\;  K_{8,11}^{J}=I_{7}^{J},\;
K_{8,12}^{J}=-N_1 I_{6}^{J},\; K_{8,13}^{J}=N_1 N_2 I_{5}^{J},\nonumber  \\
& & K_{8,14}^{J}=-N_1 I_{4}^{J},\;K_{8,15}^{J}=I_{3}^{J},\;
K_{8,16}^{J}=-N_2 I_{2}^{J};\nonumber  \\
& & K_{9,1}^{J}=I_{9}^{J},\;K_{9,2}^{J}=-N_2 I_{8}^{J},\;
K_{9,3}^{J}=I_{7}^{J},\;K_{9,4}^{J}=-N_1 I_{6}^{J},\;
K_{9,5}^{J}=N_1 N_2 I_{5}^{J},\;K_{9,6}^{J}=-N_1 I_{4}^{J},\nonumber  \\
& & K_{9,7}^{J}=I_{3}^{J},\;K_{9,8}^{J}=-N_2 I_{2}^{J},\;
K_{9,9}^{J}=I_{1}^{J};\nonumber  \\
& & K_{10,2}^{J}=-I_{2}^{J},\;K_{10,8}^{J}=-I_{8}^{J},\;
K_{10,10}^{J}=-N_2 I_{10}^{J},\;K_{10,11}^{J}=-I_{11}^{J},\;
K_{10,12}^{J}=-N_1 I_{12}^{J},\;\nonumber  \\ & &
K_{10,13}^{J}=-N_1 N_2 I_{13}^{J},\;
K_{10,14}^{J}=-N_1 I_{14}^{J},\;K_{10,15}^{J}=-I_{15}^{J},\;
K_{10,16}^{J}=-N_2 I_{16}^{J};\nonumber  \\
& & K_{11,3}^{J}=-I_{2}^{J},\;K_{11,7}^{J}=I_{8}^{J},\;
K_{11,11}^{J}=-N_2 I_{10}^{J},\;K_{11,15}^{J}=N_2 I_{16}^{J},\;
K_{11,17}^{J}=I_{11}^{J},\;K_{11,18}^{J}=N_1 I_{12}^{J},\nonumber  \\
& & K_{11,19}^{J}=N_1 N_2 I_{13}^{J},\;K_{11,20}^{J}=N_1 I_{14}^{J},\;
K_{11,21}^{J}=I_{15}^{J};\nonumber  \\
& & K_{12,4}^{J}=-I_{2}^{J},\;K_{12,6}^{J}=-I_{8}^{J},\;
K_{12,12}^{J}=-N_2 I_{10}^{J},\;K_{12,14}^{J}=-N_2 I_{16}^{J},\;
K_{12,18}^{J}=I_{11}^{J},\;K_{12,20}^{J}=-I_{15}^{J},\nonumber  \\
& & K_{12,22}^{J}=-N_1 I_{12}^{J},\;K_{12,23}^{J}=-N_1 N_2 I_{13}^{J},\;
K_{12,24}^{J}=-N_1 I_{14}^{J};\nonumber  \\
& & K_{13,5}^{J}=I_{8}^{J}-I_{2}^{J},\;
K_{13,13}^{J}=N_2 (I_{16}^{J}-I_{10}^{J}),\;
K_{13,19}^{J}=I_{11}^{J}+I_{15}^{J},\;
K_{13,23}^{J}=N_1 (I_{14}^{J}-I_{12}^{J}),\nonumber  \\
& & K_{13,25}^{J}=N_1 N_2 I_{13}^{J};\nonumber  \\
& & K_{14,4}^{J}=-I_{8}^{J},\;K_{14,6}^{J}=-I_{2}^{J},\;
K_{14,12}^{J}=-N_2 I_{16}^{J},\;K_{14,14}^{J}=-N_2 I_{10}^{J},\;
K_{14,18}^{J}=-I_{15}^{J},\;K_{14,20}^{J}=I_{11}^{J},\nonumber  \\
& & K_{14,22}^{J}=-N_1 I_{14}^{J},\;K_{14,23}^{J}=N_1 N_2 I_{13}^{J},\;
K_{14,24}^{J}=-N_1 I_{12}^{J};\nonumber  \\
& & K_{15,3}^{J}=I_{8}^{J},\;K_{15,7}^{J}=-I_{2}^{J},\;
K_{15,11}^{J}=N_2 I_{16}^{J},\;K_{15,15}^{J}=-N_2 I_{10}^{J},\;
K_{15,17}^{J}=I_{15}^{J},\;K_{15,18}^{J}=-N_1 I_{14}^{J},\nonumber  \\
& & K_{15,19}^{J}=N_1 N_2 I_{13}^{J},\;K_{15,20}^{J}=
-N_1 I_{12}^{J},\; K_{15,21}^{J}=I_{11}^{J};\nonumber  \\
& & K_{16,2}^{J}=-I_{8}^{J},\;K_{16,8}^{J}=-I_{2}^{J},\;
K_{16,10}^{J}=-N_2 I_{16}^{J},\;K_{16,11}^{J}=I_{15}^{J},\;
K_{16,12}^{J}=-N_1 I_{14}^{J},\;K_{16,13}^{J}=N_1 N_2 I_{13}^{J},\nonumber  \\
& & K_{16,14}^{J}=-N_1 I_{12}^{J},\;K_{16,15}^{J}=
I_{11}^{J},\; K_{16,16}^{J}=-N_2 I_{10}^{J};\nonumber  \\
& & K_{17,3}^{J}=I_{3}^{J},\;K_{17,7}^{J}=I_{7}^{J},\;
K_{17,11}^{J}=N_2 I_{11}^{J},\;K_{17,15}^{J}=N_2 I_{15}^{J},\;
K_{17,17}^{J}=I_{17}^{J},\;K_{17,18}^{J}=N_1 I_{18}^{J},\nonumber  \\
& & K_{17,19}^{J}=N_1 N_2 I_{19}^{J},\;K_{17,20}^{J}=
N_1 I_{20}^{J},\; K_{17,21}^{J}=I_{21}^{J};\nonumber  \\
& & K_{18,4}^{J}=I_{3}^{J},\;K_{18,6}^{J}=-I_{7}^{J},\;
K_{18,12}^{J}=N_2 I_{11}^{J},\;K_{18,14}^{J}=-N_2 I_{15}^{J},\;
K_{18,18}^{J}=I_{17}^{J},\;K_{18,20}^{J}=-I_{21}^{J},\nonumber  \\
& & K_{18,22}^{J}=-N_1 I_{18}^{J},\;K_{18,23}^{J}=
-N_1 N_2 I_{19}^{J},\; K_{18,24}^{J}=-N_1 I_{20}^{J};\nonumber  \\
& & K_{19,5}^{J}=I_{3}^{J}+I_{7}^{J}, \;
K_{19,13}^{J}=N_2 ( I_{11}^{J}+I_{15}^{J}),\;
K_{19,19}^{J}=I_{17}^{J}+I_{21}^{J},\;
K_{19,23}^{J}=N_1 ( I_{20}^{J}-I_{18}^{J}),\;\nonumber  \\ & &
K_{19,25}^{J}=N_1 N_2 I_{19}^{J};\nonumber  \\
& & K_{20,4}^{J}=-I_{7}^{J},\;K_{20,6}^{J}=I_{3}^{J},\;
K_{20,12}^{J}=-N_2 I_{15}^{J},\;K_{20,14}^{J}=N_2 I_{11}^{J},\;
K_{20,18}^{J}=-I_{21}^{J},\;K_{20,20}^{J}=I_{17}^{J},\nonumber  \\
& & K_{20,22}^{J}=-N_1 I_{20}^{J},\;K_{20,23}^{J}=
N_1 N_2 I_{19}^{J},\; K_{20,24}^{J}=-N_1 I_{18}^{J};\nonumber\\
& & K_{21,3}^{J}=I_{7}^{J},\;K_{21,7}^{J}=I_{3}^{J},\;
K_{21,11}^{J}=N_2 I_{15}^{J},\;K_{21,15}^{J}=N_2 I_{11}^{J},\;
K_{21,17}^{J}=I_{21}^{J},\;K_{21,18}^{J}=-N_1 I_{20}^{J},\nonumber  \\
& & K_{21,19}^{J}=N_1 N_2 I_{19}^{J},\;K_{21,20}^{J}=
-N_1 I_{18}^{J},\; K_{21,21}^{J}=I_{17}^{J};\nonumber\\
& & K_{22,4}^{J}=-I_{4}^{J},\;K_{22,6}^{J}=-I_{6}^{J},\;
K_{22,12}^{J}=-N_2 I_{12}^{J},\;K_{22,14}^{J}=-N_2 I_{14}^{J},\;
K_{22,18}^{J}=-I_{18}^{J},\;K_{22,20}^{J}=-I_{20}^{J},\nonumber  \\
& & K_{22,22}^{J}=-N_1 I_{22}^{J},\;K_{22,23}^{J}=
-N_1 N_2 I_{23}^{J},\; K_{22,24}^{J}=-N_1 I_{24}^{J};\nonumber\\
& & K_{23,5}^{J}=I_{6}^{J}-I_{4}^{J}, \;
K_{23,13}^{J}=N_2 ( I_{14}^{J}-I_{12}^{J}),\;
K_{23,19}^{J}=I_{20}^{J}-I_{18}^{J},\;
K_{23,23}^{J}=N_1 ( I_{24}^{J}-I_{22}^{J}),\;\nonumber  \\ & &
K_{23,25}^{J}=N_1 N_2 I_{23}^{J};\nonumber  \\
& & K_{24,4}^{J}=-I_{6}^{J},\;K_{24,6}^{J}=-I_{4}^{J},\;
K_{24,12}^{J}=-N_2 I_{14}^{J},\;K_{24,14}^{J}=-N_2 I_{12}^{J},\;
K_{24,18}^{J}=-I_{20}^{J},\;K_{24,20}^{J}=-I_{18}^{J},\nonumber  \\
& & K_{24,22}^{J}=-N_1 I_{24}^{J},\;K_{24,23}^{J}=
N_1 N_2 I_{23}^{J},\; K_{24,24}^{J}=-N_1 I_{22}^{J};\nonumber\\
& & K_{25,5}^{J}=2 I_{5}^{J}, \;
K_{25,13}^{J}=2 N_2 I_{13}^{J},\;
K_{25,19}^{J}=2 I_{19}^{J},\;
K_{25,23}^{J}=2 N_1 I_{23}^{J},\;
K_{25,25}^{J}=N_1 N_2 I_{25}^{J}\, .\nonumber
\end{eqnarray}
In these formulas
$\displaystyle{N_1 = \frac{|\mathbf{k}'|^{2}-{k'}_{0}^{2}}{m^2}},\;$
and
$\displaystyle{N_2 = \frac{|\mathbf{q}'|^{2}-{q'}_{0}^{2}}{m^2}} $
are the normalization coefficients for the vector particles
whose 4-momenta are
$k' = ({k'}_{0},\,\mathbf{k}')$ and
$q' = ({q'}_{0},\,\mathbf{q}'). $

\end{document}